\documentclass[%
 prd,
 reprint,
superscriptaddress,
nofootinbib,
 amsmath,amssymb,
 aps,
]{revtex4-2}

\usepackage{graphicx}% Include figure files
\usepackage{dcolumn}% Align table columns on decimal point
\usepackage{bm}% bold math
\usepackage{hyperref}% add hypertext capabilities
\usepackage{url}
\usepackage{tabularx}
\usepackage{newtxtext,newtxmath}
\usepackage{bm}
\usepackage[T1]{fontenc}
\usepackage{xcolor}
\usepackage{tcolorbox}
\usepackage{orcidlink}
\usepackage{bigints}

\newcommand{\mnras}{Mon.~Not.~Roy.~Astron.~Soc.}
\newcommand{\physrep}{Phys.~Rep.}
\newcommand{\jcap}{JCAP}
\newcommand{\pasj}{Publ.~Astron.~Soc.~Japan}

\newcommand{\apjs}{\apj~Suppl.}
\newcommand{\aap}{Astronomy~\&~Astrophysics}
\newcommand{\aj}{The~Astronomical~J.}

\begin{document}

\title{Modelling Galaxy Clustering and Tomographic Galaxy-Galaxy Lensing with HSC Y3 and SDSS using the Point-Mass Correction Model and Redshift Self-Calibration}

\author{Tianqing~Zhang\orcidlink{0000-0002-5596-198X}}
\thanks{tq.zhang@pitt.edu}
\affiliation{Department of Physics and Astronomy and PITT PACC, University of Pittsburgh, Pittsburgh, PA 15260, USA}
\affiliation{McWilliams Center for Cosmology and Astrophysics, Department of Physics, Carnegie
Mellon University, 5000 Forbes Ave, Pittsburgh, PA 15213, USA}

\author{Sunao~Sugiyama\orcidlink{0000-0003-1153-6735}}
\affiliation{Department of Physics and Astronomy, University of Pennsylvania, Philadelphia, PA 19104, USA}

\author{Surhud~More\orcidlink{0000-0002-2986-2371}}
\affiliation{The Inter-University Centre for Astronomy and Astrophysics, Post
bag 4, Ganeshkhind, Pune 411007, India}
\affiliation{Kavli Institute for the Physics and Mathematics of the Universe (WPI), The University of Tokyo Institutes for Advanced Study (UTIAS), The University of Tokyo, Chiba 277-8583, Japan}

\author{Rachel~Mandelbaum\orcidlink{0000-0003-2271-1527}}
\affiliation{McWilliams Center for Cosmology and Astrophysics, Department of Physics, Carnegie
Mellon University, 5000 Forbes Ave, Pittsburgh, PA 15213, USA}

\author{Xiangchong~Li\orcidlink{0000-0003-2880-5102}}
\affiliation{McWilliams Center for Cosmology and Astrophysics, Department of Physics, Carnegie
Mellon University, 5000 Forbes Ave, Pittsburgh, PA 15213, USA}
\affiliation{Brookhaven National Laboratory, Bldg 510, Upton, New York 11973, USA}

\author{Roohi~Dalal\orcidlink{0000-0002-7998-9899}}
\affiliation{American Astronomical Society, 1667 K St NW 800, Washington, D.C. 20006}

\author{Arun~Kannawadi\orcidlink{0000-0001-8783-6529}}
\affiliation{Department of Physics, Duke University, Durham, NC 27708, USA}

\author{Hironao~Miyatake\orcidlink{0000-0001-7964-9766}}
\affiliation{Kavli Institute for the Physics and Mathematics of the Universe (WPI), The University of Tokyo Institutes for Advanced Study (UTIAS), The University of Tokyo, Chiba 277-8583, Japan}
\affiliation{Kobayashi-Maskawa Institute for the Origin of Particles and the
Universe (KMI), Nagoya University, Nagoya, 464-8602, Japan}
\affiliation{Institute for Advanced Research, Nagoya University, Nagoya
464-8601, Japan}

\author{Atsushi~J.~Nishizawa\orcidlink{0000-0002-6109-2397}}
\affiliation{Kobayashi-Maskawa Institute for the Origin of Particles and the Universe (KMI), Nagoya University, Nagoya, 464-8602, Japan}
\affiliation{Gifu Shotoku Gakuen University, 1-1 Takakuwanishi, Yanaizu, Gifu, 501-6194, Japan}

 \author{Takahiro~Nishimichi\orcidlink{0000-0002-9664-0760}}
 \affiliation{Kavli Institute for the Physics and Mathematics of the Universe (WPI), The University of Tokyo Institutes for Advanced Study (UTIAS), The University of Tokyo, Chiba 277-8583, Japan}
 \affiliation{Center for Gravitational Physics, Yukawa Institute for Theoretical
Physics, Kyoto University, Kyoto 606-8502, Japan}
\affiliation{Department of Astrophysics and Atmospheric Sciences, Faculty of Science, Kyoto Sangyo University, Motoyama, Kamigamo, Kita-ku, Kyoto 603-8555, Japan}

\author{Masamune~Oguri\orcidlink{0000-0003-3484-399X}}
\affiliation{Kavli Institute for the Physics and Mathematics of the Universe (WPI), The University of Tokyo Institutes for Advanced Study (UTIAS), The University of Tokyo, Chiba 277-8583, Japan}
\affiliation{Center for Frontier Science, Chiba University, Chiba 263-8522, Japan}
\affiliation{Department of Physics, Graduate School of Science, Chiba University, Chiba 263-8522, Japan}

\author{Ken~Osato\orcidlink{0000-0002-7934-2569}}
\affiliation{Kavli Institute for the Physics and Mathematics of the Universe (WPI), The University of Tokyo Institutes for Advanced Study (UTIAS), The University of Tokyo, Chiba 277-8583, Japan}
\affiliation{Center for Frontier Science, Chiba University, Chiba 263-8522, Japan}
\affiliation{Department of Physics, Graduate School of Science, Chiba University, Chiba 263-8522, Japan}

\author{Markus~M.~Rau\orcidlink{0000-0003-3709-1324}}
\affiliation{School of Mathematics, Statistics and Physics,Newcastle University, Newcastle upon Tyne, NE17RU, United Kingdom}
\affiliation{High Energy Physics Division, Argonne National Laboratory, Lemont, IL 60439, USA}

\author{Masato~Shirasaki\orcidlink{0000-0002-1706-5797}}
\affiliation{National Astronomical Observatory of Japan, National Institutes of
Natural Sciences, Mitaka, Tokyo 181-8588, Japan}
\affiliation{The Institute of Statistical Mathematics, Tachikawa, Tokyo
190-8562, Japan}

\author{Tomomi~Sunayama\orcidlink{0009-0004-6387-5784}}
\affiliation{Kavli Institute for the Physics and Mathematics of the Universe (WPI), The University of Tokyo Institutes for Advanced Study (UTIAS), The University of Tokyo, Chiba 277-8583, Japan}
\affiliation{Kobayashi-Maskawa Institute for the Origin of Particles and the Universe (KMI), Nagoya University, Nagoya, 464-8602, Japan}
\affiliation{Academia Sinica Institute of Astronomy and Astrophysics (ASIAA), No.1, Sec. 4, Roosevelt Rd, Taipei 106319, Taiwan, R.O.C}

\author{Masahiro~Takada\orcidlink{0000-0002-5578-6472}}
\affiliation{Kavli Institute for the Physics and Mathematics of the Universe (WPI), The University of Tokyo Institutes for Advanced Study (UTIAS), The University of Tokyo, Chiba 277-8583, Japan}
\affiliation{Center for Data-Driven Discovery (CD3), Kavli IPMU (WPI), UTIAS, The University of Tokyo, Kashiwa, Chiba 277-8583, Japan}

\date{\today}

\begin{abstract}
The combination of galaxy-galaxy weak lensing and galaxy clustering is a powerful probe of the cosmological model, and exploration of how to best model and extract this information from the signals is essential.
We present the measurement of the galaxy-galaxy weak lensing signals using the SDSS DR11 spectroscopic galaxies as lens galaxies, and the HSC Y3 shear catalog as source galaxies, binned into four tomographic bins by their photometric redshift. The SDSS DR11 galaxies, with a redshift range $0.15<z<0.7$, are binned into three redshift bins, each as a probe for measuring the projected correlation function, $w_p(R_p)$. We measure the galaxy-galaxy lensing signal $\Delta \Sigma (R_p)$ in 12 lens-source bin pairs and show that there is no evidence for significant systematic biases in the measurement with null testing. We combine our $w_p(R_p)$ and $\Delta \Sigma (R_p)$ ($2\times2$pt) data vectors and perform likelihood inference with a flat $\Lambda$CDM model. For $\Delta \Sigma (R_p)$, we extend the lower limit of the scale cut compared to previous HSC Y3 analyses to $2 h^{-1}$Mpc by including a point-mass correction term in addition to the minimal bias model.  We present various tests to validate our model and provide extended consistency tests. In the $\Lambda$CDM context, our fiducial model yields $S_8 = 0.804^{+0.051}_{-0.051}$. The $2\times2$pt data vector provides redshift parameter constraints for the third and fourth redshift bins $\Delta z_3 = -0.079^{+0.074}_{-0.084}$, and $\Delta z_4 = -0.203^{+0.167}_{-0.206}$, which is consistent with results from the previous tomographic cosmic shear studies, and serves as the foundation for a future $3\times 2$pt analysis. 
% \rachel{Have you considered a single sentence with the estimated $\Delta z$ shift and the improved SNR of your analysis compared to a single high-redshift bin?  The title highlights `Source redshift constraints' so it seems to make sense to put numbers here?} \tianqing{done.}

\end{abstract}

\maketitle

\section{\label{sec:introduction} Introduction}

% \rachel{Consider putting a full corner plot (all parameters) in an appendix.}\tianqing{done. }

% the success of LCDM and the challenge
In the current era of precision cosmology, the concordance $\Lambda$CDM model has received significant attention due to its success in describing a wide range of cosmological observations \cite{Peebles2025}. These include the clustering of galaxies and the weak gravitational lensing signal tracing the large-scale structure of the universe \cite{DESY3_3x2_2022, Sugiyama2023, wright2025}, Type Ia supernovae \cite{Riess2019, Scolnic2018}, and the cosmic microwave background (CMB) anisotropies \cite{Planck2018Cosmology}. Despite its successes, $\Lambda$CDM faces several challenges. Most notably, the fundamental physical nature of dark matter and dark energy remains unknown. Furthermore, tensions have emerged between the parameter constraints inferred from different cosmological probes. A prominent example is the discrepancy in the measurement of the Hubble constant, $H_0$, between local observations using Type Ia supernovae and other local probes compared to $H_0$ from CMB-based inference \cite{Hu2023}. Another emerging tension involves the amplitude of matter fluctuations, typically quantified by $\sigma_8$, or the derived parameter $S_8 = \sigma_8 \sqrt{\Omega_m/0.3}$, where weak lensing measurements suggest lower values compared to CMB predictions \cite{s8_tension2021}. These discrepancies motivate the development of next-generation surveys to map larger volumes of the universe and the refinement of methodologies to ensure unbiased data processing and analysis, both of which are essential for advancing our understanding of $\Lambda$CDM and the physics of the late-time universe.

% galaxy clustering and weak lensing as a promising cosmological probe

Under the $\Lambda$CDM paradigm, the gravitational attraction of matter (including dark matter) drives the growth of cosmic structure, while dark energy counteracts this growth by accelerating the expansion of the universe \cite{Dodelson2020}. The resulting matter distribution features over-dense regions, known as dark matter halos and cosmic web, and under-dense regions, known as cosmic voids, which are statistically characterized by the matter power spectrum, $P(k, z)$—a function of wavenumber $k$ and redshift $z$. Galaxies form within dark matter halos, making them biased tracers of the underlying matter distribution \cite{Cooray2002}. The spatial distribution of galaxies is probed through galaxy clustering measurements (e.g., \cite{DESI_KP5,DESY3_clustering}). Additionally, all matter including dark matter induces weak gravitational lensing distortions on background galaxies along the line of sight. When the lensing signal is measured around foreground galaxies, it is referred to as galaxy-galaxy lensing. Because galaxy clustering and galaxy-galaxy lensing depend differently on the galaxy bias, combining the two measurements—an approach known as the 2×2pt analysis—is a common strategy to break the degeneracy between galaxy bias and the matter power spectrum \cite{Miyatake2022}. Furthermore, the 2×2pt analysis is often combined with cosmic shear measurements, resulting in the so-called 3×2pt analysis, which provides tighter constraints on cosmological parameters.

% HSC survey

The Hyper Suprime-Cam (HSC; \cite{Aihara2018, Miyazaki2018}) is a wide-field imaging survey camera mounted at the prime focus of the 8.2-meter Subaru Telescope. The wide field of view (1.77 deg$^2$), combined with the excellent seeing conditions on Maunakea, makes HSC a highly effective instruments for large-volume imaging surveys. The HSC Wide survey will provide multi-band imaging over an area of $1100$ deg$^2$ upon completion, creating an exquisite dataset for weak lensing measurements. In this paper, we use the HSC Year-3 shear catalog \cite{Li2022}, which covers $433$ deg$^2$ across six HSC fields, to measure the galaxy-galaxy lensing signal. For the lens sample, we use galaxies from Data Release 11 (DR11) of the Sloan Digital Sky Survey (SDSS) Baryon Oscillation Spectroscopic Survey (BOSS) \cite{sdss_dr11}, which trace the large-scale structure and serve as the lenses for the lensing measurement.

% challenge realized in the last HSC Y3 cosmological analysis campaign, how this study is designed to answer that question

In the initial HSC Y3 cosmological analyses, two cosmic shear studies using four tomographic redshift bins were conducted: one in real space \cite{Li2023} and one in Fourier space \cite{Dalal2023}. 3×2pt analyses combining a single HSC source bin with SDSS DR11 lens galaxies were also performed \cite{More2023, Sugiyama2023, Miyatake2023}, due to their effectiveness in extracting the galaxy-galaxy lensing signal while reducing and/or self-calibrating several sources of uncertainty with relatively shorter data vectors \cite{oguri2011}. A critical challenge emerged during this analysis campaign: the mean redshifts of tomographic bins 3 and 4, as provided by the photometric redshift estimates \cite{Rau2022}, appeared inconsistent with the mean redshifts preferred by $\Lambda$CDM fits. To ensure that the cosmological results remained unbiased, the previous HSC Y3 analyses adopted uninformative priors on the mean redshifts of tomographic bins 3 and 4, which consequently reduced their constraining power on cosmological parameters.

% The idea of shear ratio 

One way to calibrate the mean redshift is to leverage the ``shear ratio'' method in galaxy-galaxy lensing \cite{Prat2018}. This technique estimates the mean redshift of a galaxy ensemble by comparing the amplitudes of the lensing signals from lens bins at different redshifts, relying purely on geometric considerations under the assumption of a cosmological model. By incorporating galaxy-galaxy lensing into the likelihood analysis, we can implicitly use the shear ratio to constrain the mean redshift more tightly than the prior, a process often referred to as ``self-calibration'' \cite{Sugiyama2023}.

% purpose of this study and setting up for future

In this work, we perform a galaxy-galaxy lensing measurement using the SDSS DR11 and HSC Y3 shear catalog, with a tomographic redshift binning that is consistent with the binning in the HSC Y3 cosmic shear analysis. We combine this with the galaxy clustering signal. By binning the galaxy-galaxy lensing consistently with the cosmic shear analysis, we can (a) demonstrate the self-calibration power of galaxy-galaxy lensing within a 2×2pt analysis, (b) build the foundation for a 3×2pt analysis that utilizes the tomographic source samples, which offers better constraining power than single-bin analysis due to the additional information, and (c) assess the (in)consistency of the mean redshift constraints between cosmic shear and galaxy-galaxy lensing self-calibration.

This work also explores methodologies to combine small-scale and large-scale information in galaxy-galaxy lensing to maximize the overall signal-to-noise ratio. By adding a point-mass correction term \citep{Baldauf2010,Mandelbaum2013} to $\Delta \Sigma(R_p)$ modeled by the linear bias model \cite{Sugiyama2022}, we can push the lower limit of the scale cut for $\Delta \Sigma(R_p)$ from $8\ h^{-1}\mathrm{Mpc}$ to $2\ h^{-1}\mathrm{Mpc}$. We use mock data vectors from realistic $N$-body simulations \cite{Nishimichi2019} to ensure that the result is not dominated by systematic errors.

The cosmological analysis performed in this work combines the galaxy clustering signal, $w_p(R_p)$, measured in \cite{More2023}, with the tomographic galaxy-galaxy lensing signal, $\Delta \Sigma(R_p)$. The lensing signal measurements, validation tests, and consistency checks are all carried out under catalog-level blinding, which protects the analysis from confirmation bias. 
% Our fiducial model yields $S_8 = 0.804^{+0.051}_{-0.051}$ and mean redshift biases for the third and fourth bins of the HSC Y3 shear catalog of $\Delta z_3 = -0.079^{+0.074}_{-0.084}$, and $\Delta z_3 = -0.203^{+0.167}_{-0.206}$, respectively, which are statistically consistent with the results from \cite{Dalal2023, Li2023}.

The paper is organized as follows. Section~\ref{sec:data:0} describes the data used in this work, including the HSC Y3 shear catalog, the SDSS DR11 galaxies with spectroscopic redshift measurements, and the mock catalog that mimics the HSC and SDSS catalogs. Section~\ref{sec:measurement:0} details the measurement of the galaxy-galaxy lensing signal and the null tests used to validate the measurements. Section~\ref{sec:model:0} outlines the cosmological model used to describe the galaxy clustering and galaxy-galaxy lensing data vectors, as well as the framework of the likelihood analysis.
Section~\ref{sec:validation:0} describes the validation tests and internal consistency tests performed in this work. 
Section~\ref{sec:results:0} presents the results of the baseline model, validation tests, and consistency tests. Finally, Section~\ref{sec:conclusion:0} summarizes our findings and discusses future developments.

\section{\label{sec:data:0} Data}

\begin{figure}
\includegraphics[width=0.95\columnwidth]{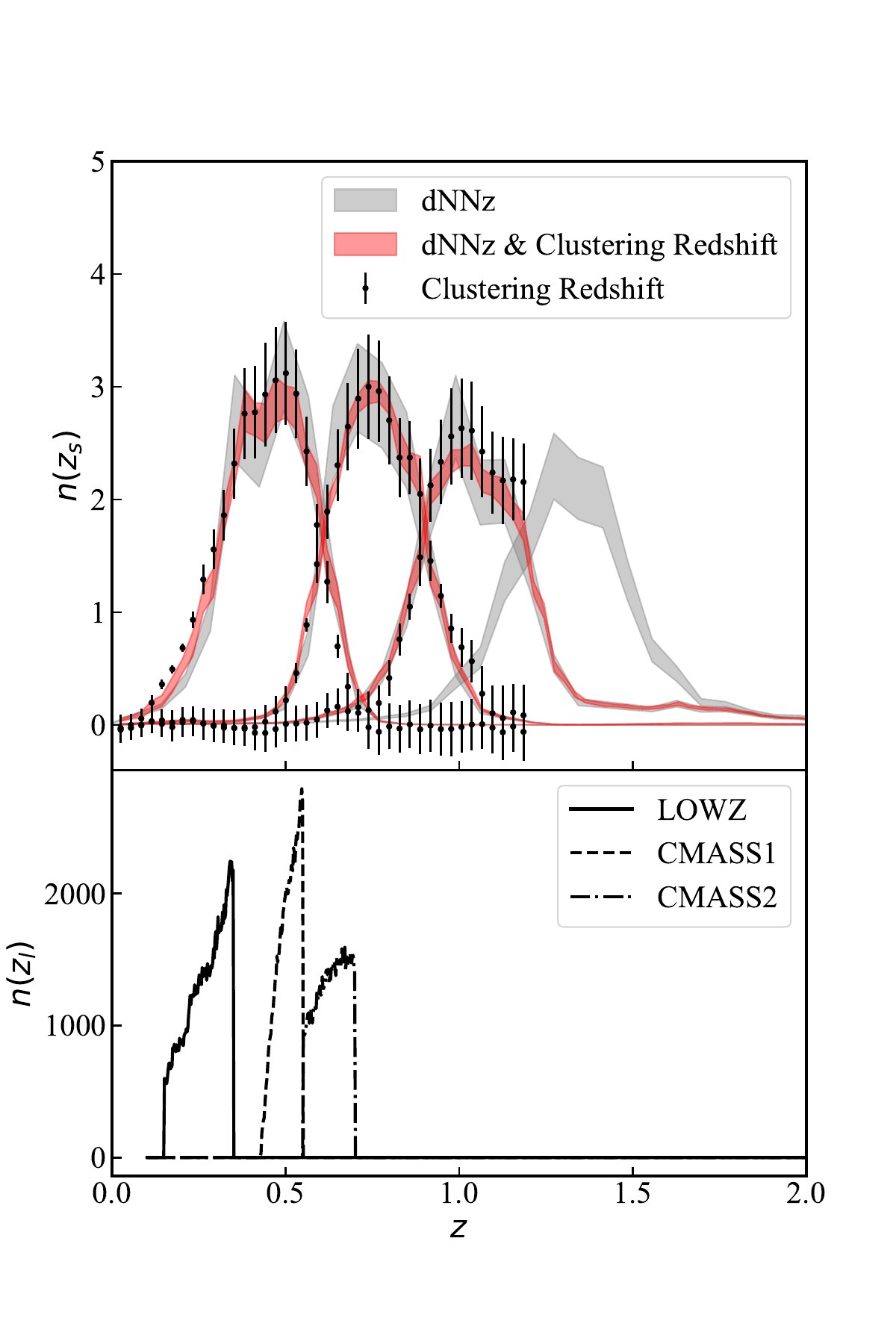}
\caption{\label{fig:redshift_distro} The redshift distributions of the HSC Y3 source galaxies (upper panel) and the SDSS lens galaxies (lower panel) are shown. The first and second source redshift distributions are fully calibrated through cross-correlation with the CAMIRA-LRG samples within the HSC Y3 footprint, while the third bin is only partially calibrated and the fourth bin remains uncalibrated. }
\end{figure}

%structure of this section

This section describes the data products used in this work. To conduct the joint analysis of galaxy clustering and galaxy-galaxy lensing, both a lens catalog and a source catalog are required. The lens catalog is the large-scale structure sample from Data Release 11 (DR11) \cite{sdss_dr11} of SDSS-III's Baryon Oscillation Spectroscopic Survey (BOSS; \cite{boss2013}), described in Section~\ref{sec:data:sdss}. The source catalog is the HSC Y3 shear catalog, described in Section~\ref{sec:data:hscy3}.  The blinding process is performed at the catalog level, as described in Section~\ref{sec:data:blinding}. To measure the covariance matrix of the galaxy-galaxy lensing signal, we use a mock catalog with realistically simulated lens and source galaxies, described in Section~\ref{sec:data:mock}.

\subsection{\label{sec:data:sdss} SDSS Catalog}

We use galaxies from the SDSS-III survey with spectroscopic redshift measurements to trace the large-scale structure responsible for the lensing effect. Specifically, our lens catalog is the BOSS large-scale structure sample from Data Release 11 (DR11) \cite{sdss_dr11} of SDSS. This galaxy sample was previously used in the 2×2pt analysis of the first-year HSC data \cite{Miyatake2022, Sugiyama2020} and in the HSC Y3 3×2pt analysis \cite{More2023, Sugiyama2023, Miyatake2023}. Here, we provide only a concise description of the catalog and refer readers to the BOSS papers and the earlier HSC papers for detailed information.

The Baryon Oscillation Spectroscopic Survey (BOSS) is a spectroscopic follow-up survey targeting galaxies and quasars selected from SDSS-I/II imaging, covering approximately 11,000 deg$^2$ of the sky \cite{ABAZAJIAN2009}. The survey was conducted using the 2.5-meter SDSS telescope \cite{gunn2006}. The BOSS large-scale structure samples include galaxies across a broad redshift range: $0.15 < z_l < 0.35$ for the LOWZ sample and $0.43 < z_l < 0.70$ for the CMASS sample. The original BOSS galaxy sample over $z \in [0.15, 0.70]$ is neither volume-limited nor flux-limited, and can cause systematic bias on galaxy clustering measurement \cite{hang2024}. To create an approximately volume-limited sample, we define three subsamples by applying cuts on the absolute magnitude: $M_i - 5 \log h < -21.5$, $-21.9$, and $-22.2$ for LOWZ galaxies with $0.15 < z_l < 0.35$, CMASS1 galaxies with $0.43 < z_l < 0.55$, and CMASS2 galaxies with $0.55 < z_l < 0.70$, respectively. 
The resulting subsamples have volumetric densities of 1.8, 0.74, and 0.45 $\times 10^{-4}\ (h^{-1}\ \mathrm{Mpc})^{-3}$, respectively. 
The redshift distributions of the lens samples are shown in the bottom panel of Fig.~\ref{fig:redshift_distro}.

In Fig.~\ref{fig:redshift_distro}, we see that our lens and source galaxies have significant overlap in redshift space. In Section~\ref{sec:measurement:ggl}, we show that by applying optimal weighting to the lens-source pairs based on the source photometric redshift, we minimize the dilution effect caused by the overlapping redshift distributions of sources and lenses. However, due to photometric redshift uncertainties, we expect that some source galaxies with nonzero weights will be physically associated with the lens galaxies. We detect this effect through the boost factor, as explained in Section~\ref{sec:measurement:null}, and mitigate its impact by multiplying the galaxy-galaxy lensing signal by the corresponding boost factor.
For the lens-source galaxies pairs where the source galaxies are in front of the lens galaxies, they are given zero weights, as explained in Section~\ref{sec:measurement:ggl}. 
The measurement of the boost factor involves a set of random points generated within the BOSS footprint. These random points are constructed by downsampling the DR11 galaxies such that they follow the redshift distribution of the LOWZ and CMASS samples. The random catalogs are used both to measure the boost factor and to estimate the systematic and statistical errors of the galaxy-galaxy lensing and clustering signals. The systematic signal measured around random lenses is subtracted from the signal measured around real lenses, as will be described in detail in Section~\ref{sec:measurement:ggl}.

It has been shown in \cite{More2023} that the lens galaxy bias of the clustering and galaxy-galaxy lensing signals do not evolve significantly within the redshift bins, due to the approximately volume-limited nature of the lens sample. Since we use the same lens sample in this work, we do not repeat that test here.

\subsection{\label{sec:data:hscy3} HSC-Y3 Shear Catalog}

% general info about the HSC survey and data product

The Hyper Suprime-Cam (HSC) \cite{Furusawa2018, Miyazaki2018} is a wide-field optical camera mounted at the prime focus of the 8.2-meter Subaru Telescope, located at Maunakea, Hawaii. The HSC Subaru Strategic Program (HSC-SSP) \cite{Aihara2018} is a deep, multiband imaging survey targeting a footprint of 1400 deg$^2$ upon completion. The Wide layer of HSC surveys the sky in five bands, \textit{grizy}. Galaxy shape measurements are performed in the \textit{i}-band, which reaches a $5\sigma$ detection threshold at $i \sim 26$. The median seeing in the \textit{i}-band images is approximately $0.6$ arcsec.

% general info about the three year shape catalog

The three-year shape catalog \cite{Li2022}\footnote{The shape catalog is publicly available as a catalog of the PDR3 \cite{Aihara2022} at \url{https://hsc-release.mtk.nao.ac.jp/doc/index.php/data-access\_\_pdr3/}.} used as the weak lensing source shear catalog in this work was produced from an internal data release (S19A), between the second and third public data releases \cite{Aihara2019, Aihara2022}. 
The catalog was processed using hscPipe v7 \cite{Nishizawa2020}, a fork of the Rubin Observatory's LSST Science Pipeline\footnote{\url{https://pipelines.lsst.io/}}, with several improvements over the earlier data releases processed with the original HSC image processing pipeline \cite{Bosch2018}. For a detailed description of the catalog, we refer readers to \cite{Li2022}. In brief, the HSC Y3 shear catalog consists of galaxies in the full-depth, full-color regions observed in five HSC filters: \textit{grizy}. The selection criteria include model magnitude $i < 24.5$, $i$-band flux signal-to-noise ratio (SNR) > 10, resolution factor $R_2 > 0.3$, estimated shape noise < 0.4, detection SNR > 5 in at least two bands other than $i$, aperture magnitude in $i$-band < 25.5, and $i$-band blendedness < $10^{-0.38}$.

The shape catalog contains a total of 35.7 million galaxies, covering a footprint of 433 deg$^2$. The effective source number density is 19.9 arcmin$^{-2}$. The HSC footprint is divided into six fields: GAMA09H, GAMA15H, WIDE12H, XMM, VVDS, and HECTOMAP. Galaxy shapes in the catalog are measured using the Re-Gaussianization method \cite{Hirata2003, Mandelbaum2005}, as implemented in GalSim \cite{Rowe2014}. The multiplicative shear calibration is derived from image simulations matched to the properties of the HSC catalog, ensuring a multiplicative shear bias below $10^{-2}$. Various systematic tests and null tests are presented to confirm the absence of significant additive shear biases.
Throughout this work, we exclude a patch of the sky in the GAMA09H region due to the presence of an excessive B-mode signal found in the cosmic shear analysis \cite{Li2023}. The area of the excluded patch is about 20 deg$^2$, reducing the total area used in this analysis to 416 deg$^2$.

The redshifts of the galaxies in the HSC Y3 shape catalog are estimated using three photometric redshift (photo-$z$) algorithms \cite{Nishizawa2020}. \textsc{Mizuki} is a template-fitting-based photo-$z$ code, while \textsc{DNNz} and \textsc{DEmPz} are two machine-learning-based methods. All three methods provide a probabilistic density function (PDF), $P(z_s)$, for each galaxy in the catalog.

The source galaxies are split into four tomographic bins following the binning method used in \cite{Li2023}, assigning galaxies to redshift bins 1–4 with the ranges [0.3–0.6], [0.6–0.9], [0.9–1.2], and [1.2–1.5], based on their $z_{\text{best}}$ from \textsc{DNNz}. In addition, we find that $\sim$31\% of the galaxies in the first bin and $\sim$8\% of the galaxies in the second bin exhibit double peaks in their \textsc{Mizuki} and \textsc{DNNz} PDFs. To remove these galaxies, we apply the following selection criteria:
\begin{equation} 
\left(z_{\text{0.975}; i}^\text{mizuki} - z_{\text{0.025}; i}^\text{mizuki}\right) < 2.7 \quad \text{and} \quad \left(z_{\text{0.975}; i}^\text{dnnz} - z_{\text{0.025}; i}^\text{dnnz}\right) < 2.7, \label{eq:selection} 
\end{equation} 
where $z_{\text{0.975}; i}^{\text{mizuki}(\text{dnnz})}$ and $z_{\text{0.025}; i}^{\text{mizuki}(\text{dnnz})}$ denote the 97.5th and 2.5th percentiles of the photo-$z$ PDF for galaxy $i$ from \textsc{Mizuki} (\textsc{DNNz}), respectively.

In the cosmological analysis, the galaxy redshift distribution, $n(z_s)$, is required to compute the angular power spectrum from the three-dimensional matter power spectrum. The $n(z_s)$ of the four tomographic bins combines stacked information from the photo-$z$ PDFs and cross-correlation measurements with the CAMIRA luminous red galaxies (CAMIRA-LRG) in the HSC fields \cite{Oguri2014, Oguri2018}. It is important to note that the CAMIRA-LRG sample covers the redshift range $0.1 < z < 1.1$, meaning that only the first and second tomographic bins are fully calibrated by cross-correlation. The third bin is only partially calibrated, while the fourth bin remains uncalibrated. The effective number densities in bins 1–4 are 3.77, 5.07, 4.00, and 2.12 arcmin$^{-2}$, respectively, resulting in a total of 14.96 arcmin$^{-2}$ (corresponding to a 25.1\% loss compared to the full catalog). Notably, we nearly triple the number of galaxies used in this galaxy-galaxy lensing analysis compared to the first HSC Y3 galaxy-galaxy lensing analyses \cite{More2023}, which defined the source sample as galaxies with redshift probability $P(z>0.75) > 0.99$.
The estimated redshift distributions $n(z_s)$ of the HSC Y3 catalog are shown in the top panel of Fig.~\ref{fig:redshift_distro}. The grey regions represent the posterior from the \textsc{DNNz} photo-$z$ estimates, while the red regions show the posterior after probabilistically combining the \textsc{DNNz} photo-$z$ estimates and the clustering redshift information using a logistic Gaussian process. For more details on the redshift distribution of the HSC Y3 shape catalog, we refer readers to \cite{Rau2022}.

\subsection{\label{sec:data:blinding} Blinding Strategy}

The galaxy-galaxy lensing data vector, divided into 12 lens-source bin pairs, represents a new dataset that has not been previously studied. Therefore, it is necessary to shield the analysis from confirmation bias. We follow the blinding method described in \cite{Li2022}, applying a multiplicative shear bias to blind the shear catalog. Specifically, we produce three catalogs with different multiplicative biases. For blind catalog $i = 0, 1, 2$, and for galaxy ID $j$, the multiplicative bias $m^{ij}$ is defined as \begin{equation} \label{eq:mbias_blinding} m^{ij} = m^{j} + dm_1^i + dm_2^i, \end{equation} where $m^{j}$ is the true multiplicative bias, which remains unknown to the analysis lead before unblinding. The term $dm_1^i$ is a known offset, allowing the analysis lead to compare multiple versions of the blinded catalogs without inadvertently unblinding the data. The term $dm_2^i$ for $i=0,1,2$ is randomly assigned by one of three sets of permutations: $(-0.1, -0.05, 0.0)$, $(-0.05, 0.0, 0.05)$, or $(0.0, 0.05, 0.1)$. The amplitude of $0.05$ is chosen because it matches the typical statistical uncertainty on the overall shear amplitude; setting the amplitude too large would allow the true catalog to be easily identified, while setting it too small would still permit confirmation bias to influence analysis decisions.

The galaxy-galaxy lensing measurements and the cosmological analyses are performed independently on all three blinded catalogs in parallel. Only after all null tests, validation tests, and internal consistency checks are successfully passed for all three blinded versions—and after the analysis code undergoes an internal code review—is the lead analyzer permitted to unblind and reveal the identity of the true catalog.

% A paragraph describing which one is the real catalog in the three catalog after unblinding. 

After validating the model with the mock validation analyses (Section~\ref{sec:results:validation}) and completing the internal consistency tests (Section~\ref{sec:results:internal_consistency}) for all three blinded catalogs, we proceeded to unblind the data. The \texttt{blinded\_id=0} catalog was revealed to be the true catalog. In this work, we present results only from the \texttt{blinded\_id=0} catalog.

\subsection{\label{sec:data:mock} Mock Catalogs}

\begin{figure*}
\includegraphics[width=2.0\columnwidth]{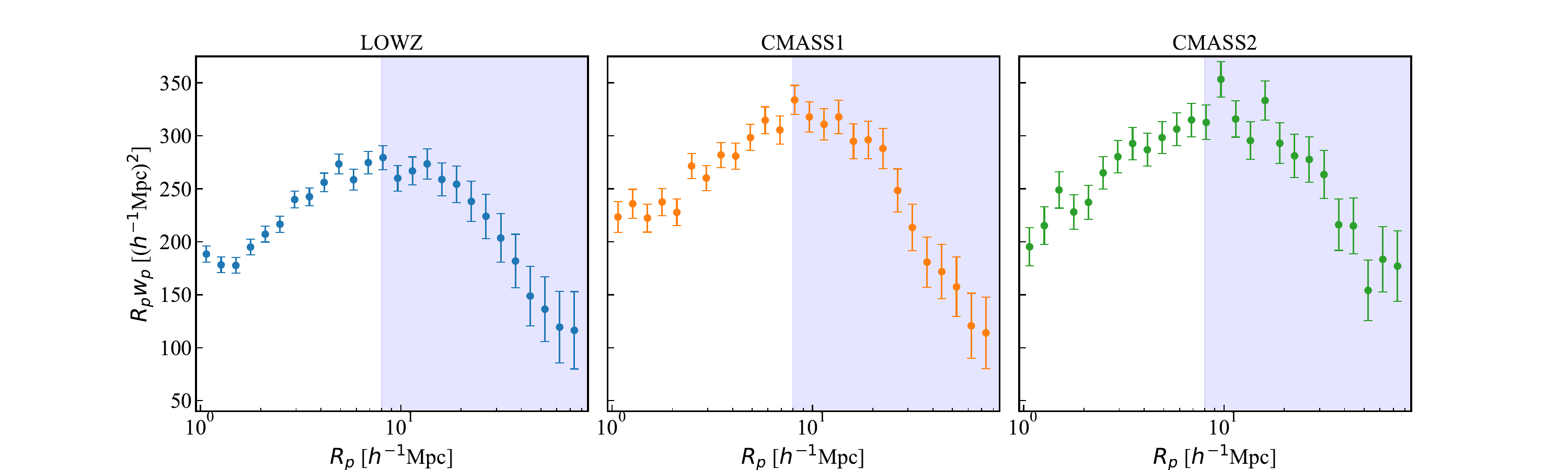}
\includegraphics[width=2.0\columnwidth]{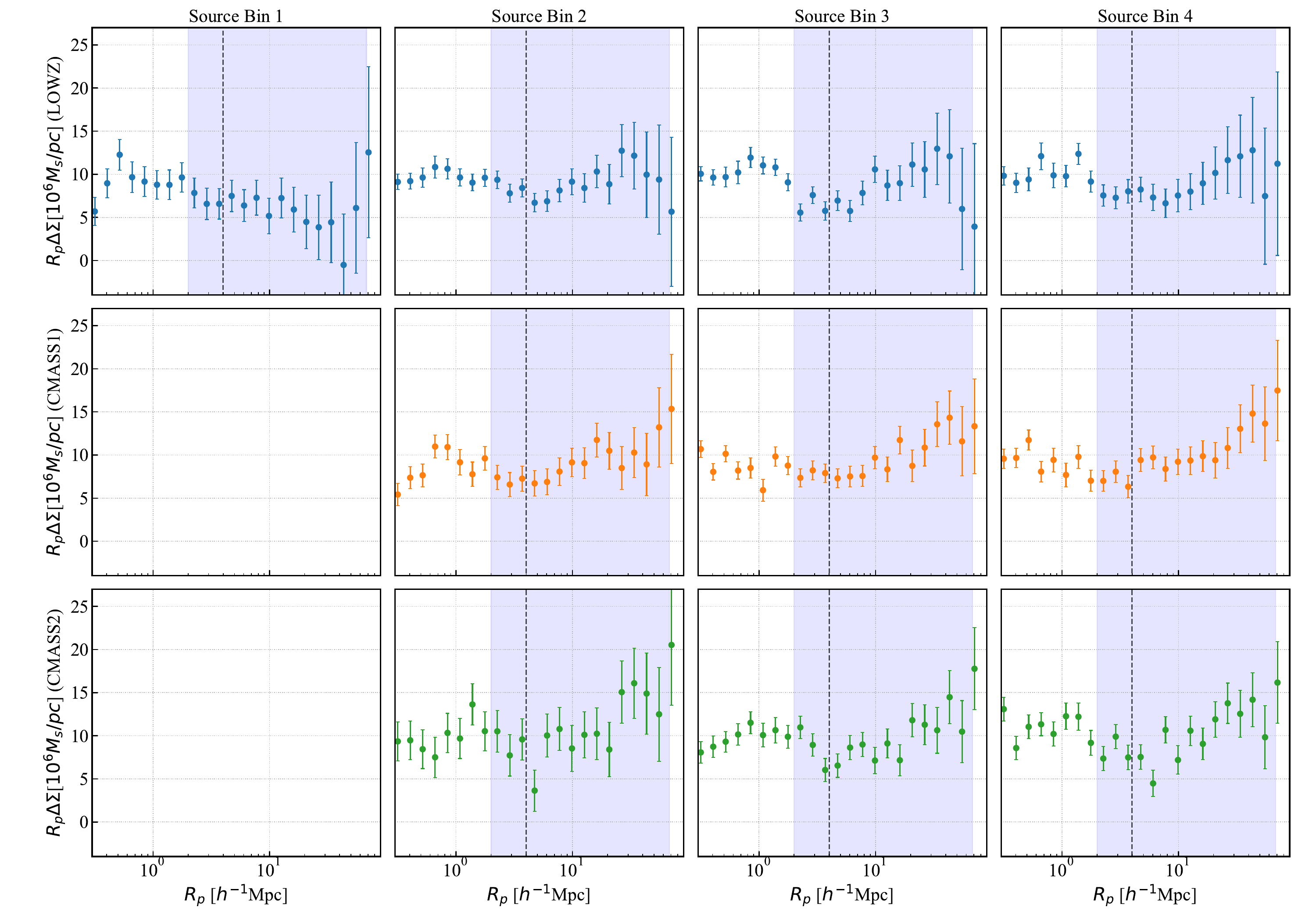}
\caption{\label{fig:measurement} The top row shows the measurement of the $w_p$ signal as a function of projected radius for the three BOSS tomographic bins, while the bottom three rows present the measurements of $\Delta \Sigma$ for the same BOSS bins, each paired with the four tomographic HSC source bins. The blue shaded regions indicate the scales used in this work for cosmological inference. The dashed lines indicate $R_0 = 4 h^{-1}{\rm Mpc}$, as introduced in Section~\ref{sec:model:0}. The color of $w_p$ and $\Delta \Sigma$ corresponds to different lens bins. %\rachel{It is a little difficult to see at a glance which panels are connected.  I would suggest using a different point color for each lens sample, so there is a visual signature of which $w_p$ and $\Delta\Sigma$ panels have a common lens sample.  Perhaps there could also then be a legend in one of the empty panels in the bottom figure?} \tianqing{Done.} \rachel{Watch for inconsistent notation in $R$ versus $R_p$.}\tianqing{\textit{Homogenized} to $R_p$}
}
\end{figure*}

A critical component of conducting Bayesian inference on the measured data vector to obtain constraints on cosmological parameters is the estimation of the covariance matrix (e.g., \cite{Singh2017}). The covariance matrix characterizes the uncertainties in the data vector. For galaxy clustering, the uncertainties arise from the sampling noise (or shot noise) due to the discrete nature of galaxies in the large-scale structure, as well as cosmic variance, which stems from the finite volume of the observed universe. For galaxy-galaxy lensing, the uncertainties originate from the intrinsic shape noise of source galaxies, the shot noise of the lens galaxies, and the cosmic variance of both the lens and source populations.
To estimate the covariance matrix for the 2×2pt analysis data vector, we use simulated galaxies to produce mock catalogs mimicking SDSS-like and HSC-like samples based on $N$-body simulations \cite{Takahashi2017}. We then measure galaxy clustering and galaxy-galaxy lensing on these catalogs to construct the covariance matrix.

For the HSC-like mock source catalogs, there are 108 full-sky simulations, each containing 13 non-overlapping HSC Y3 regions, yielding a total of 1404 realizations of the HSC mock catalogs. The dark matter halos are simulated using $N$-body simulations that adopt a flat $\Lambda$CDM cosmological model consistent with the nine-year WMAP cosmological parameters (WMAP9) \cite{WMAP9}. Convergence and shear at each source galaxy position are generated through ray-tracing.
The source galaxies in the mock catalogs have the same number, positions, and photometric redshift PDFs as those in the HSC Y3 shape catalog. The shapes of the galaxies are randomly rotated to remove any original lensing signal. The resulting intrinsic shapes in the mock catalogs are expressed as
\begin{equation}
    e^{\text{int}} = \left( \frac{e_{\rm rms}}{\sqrt{e^2_{\text{rms}} + \sigma^2_{e}}}\right) e^{\text{obs}} \exp(i\phi), 
\end{equation}
where $e_{\text{rms}}$ is the root-mean-square ellipticity of the intrinsic shapes, $\sigma_{e}^2$ is the per-object variance of the shape, $e^{\text{obs}}$ is the measured shape, and $\exp(i\phi)$ represents a random rotation applied to the measured shape, with $0 < \phi < 2\pi$. The prefactor in parentheseis ensures that the RMSrms ellipticity of the rotated intrinsic shape is equal to $e_{\rm rms}$.
The measurement error $e^{\text{mea}}$ is expressed as 
\begin{equation}
    e^{\text{mea}} = N_1 + iN_2, 
\end{equation}
where $N_1$ and $N_2$ are random numbers drawn from a normal distribution with zero mean and standard deviation $\sigma_e$.
For a detailed description of the HSC mock catalog generation, we refer readers to \cite{Shirasaki2019}.

Within the same suite of $N$-body simulations that produce the synthetic convergence and shear fields for the source galaxies, SDSS-like galaxies are populated according to a halo occupation distribution (HOD) model. The HOD defines the number of galaxies within a halo as a conditional probabilistic function of the halo mass $M$. The expected numbers of central galaxies, $\langle N_{\text{cen}} \rangle_M$, and satellite galaxies, $\langle N_{\text{sat}} \rangle_M$, are given by
\begin{align}
\label{eq:cen_halo}    \langle N_{\text{cen}} \rangle_M &= \frac{1}{2} \left[ 1+\text{erf}\left( \frac{\log_{10}M - \log_{10} M_{\text{min}}}{\sigma_{\log_{10}M}}\right) \right]\\
\label{eq:sat_halo}    \langle N_{\text{sat}} \rangle_M &=  \langle N_{\text{cen}} \rangle_M \left( \frac{M - \kappa_M M_{\text{min}}}{M_1}\right)^{\alpha_{M}}.
\end{align}
The total number of galaxies is given by $\langle N_{\text{tot}} \rangle_M = \langle N_{\text{cen}} \rangle_M + \langle N_{\text{sat}} \rangle_M$. Here, $M_{\text{min}}$ sets the minimum halo mass for hosting galaxies, and $\alpha_M$ controls the power-law slope of the satellite occupation at low mass. The HOD parameters are estimated by fitting the observed abundances of the LOWZ, CMASS1, and CMASS2 samples.
For each halo in the $N$-body simulations, we compute the number of central galaxies using Eq.~\ref{eq:cen_halo}. Central galaxies are assumed to reside at the halo center. If a halo hosts a central galaxy, we populate satellite galaxies by drawing from a Poisson distribution with mean $\lambda_M = \left( \frac{M - \kappa_M M_{\text{min}}}{M_1} \right)^{\alpha_M}$. The radial distribution of satellite galaxies follows a Navarro-Frenk-White (NFW) profile \cite{Navarro1997} with concentration parameters measured by \textsc{ROCKSTAR} \cite{behroozi2013}, a halo-finding algorithm. Satellite galaxies are assigned virial velocities with zero mean and variance $\sigma_{\text{vir}}^2 = (1+z) GM/(2R_{200m})$.

\section{\label{sec:measurement:0} Measurements}

This section describes the measurement of the data vector. We measure galaxy clustering through the autocorrelation of the LOWZ, CMASS1, and CMASS2 galaxies, and galaxy-galaxy lensing using BOSS galaxies as lenses and HSC galaxies as sources. Since the galaxy clustering measurement is identical to that used in the HSC Y3 cosmological analysis \cite{More2023}, we only briefly describe it in Section~\ref{sec:measurement:ggc}. The galaxy-galaxy lensing measurement using tomographic source bins is detailed in Section~\ref{sec:measurement:ggl}. Null tests performed in this work are described in Section~\ref{sec:measurement:null}

\subsection{\label{sec:measurement:ggc} Galaxy Clustering}

In this work, we use the galaxy clustering measurements of the BOSS galaxies from \cite{More2023}. Therefore, we only briefly describe the data vector and the measurement code here, and refer readers to \cite{More2023} for full details about the clustering measurements and associated tests.

The clustering signal is measured by the weighted pair counts as a function of projected separation $R_p$ and line-of-sight separation $\pi$, given the sky coordinates, redshift, and associated weights. The three-dimensional correlation function is estimated using the Landy–Szalay estimator $\xi(R_p,\pi)$: 
\begin{equation} 
\label{eq:LS_estimator} 
\xi(R_p,\pi) = \frac{DD - 2DR + RR}{RR}, 
\end{equation} 
where $DD$ is the number of galaxy–galaxy pairs, $DR$ is the number of galaxy–random pairs, and $RR$ is the number of random–random pairs at a given $(R_p, \pi)$. To reduce the shot noise in $DR$ and $RR$, we use a random catalog with 50 times more random points than galaxies.

The three-dimensional correlation function is integrated along the line of sight to compute the projected correlation function, $w_p(R_p)$:
\begin{equation}
\label{eq:xi2wp}
w_p (R_p) = 2 \int_0^{\pi_{\text{max}}} \xi(R_p,\pi) d\pi.
\end{equation}
Here, $\pi_{\text{max}} = 100\ h^{-1}\mathrm{Mpc}$, which is smaller than the redshift width of the subsamples and the scale of the redshift space distortions, but larger than the maximum transverse separation $R_p$. See \cite{Zehavi2005} for discussions on the impact of $\pi_{\text{max}}$ on the projected correlation function.
The conversion from redshift to comoving distance is cosmology-dependent; throughout the clustering measurements, we assume the WMAP9 cosmology \cite{WMAP9}. We note that while $w_p(R_p)$ depends on the assumed cosmology, the impact on cosmological parameter inference is negligible for small deviations around the fiducial cosmology, as discussed in \cite{more2015}.

% Describe the measurement of w_p and the scale cut

The measurement of the $w_p$ signal for separations larger than $1\ h^{-1}\mathrm{Mpc}$ is shown in Fig.~\ref{fig:measurement}. The transition from the one-halo term to the two-halo term can be observed around $1.5\ h^{-1}\mathrm{Mpc}$ for the LOWZ sample, while the transition for the CMASS1 and CMASS2 samples is less pronounced. Since we adopt the Minimal Bias model developed in Sugiyama \textit{et al.} \cite{Sugiyama2020} for modeling $w_p$, we apply the validated lower scale cut of $8\ h^{-1}\mathrm{Mpc}$.

The covariance matrix of $w_p$ is estimated using 108 realizations of full-sky $N$-body simulations with BOSS galaxies populated, as described in Section~\ref{sec:data:mock}. In each realization, the SDSS footprint is divided into 192 jackknife regions. A covariance matrix is calculated for each realization based on the jackknife resampling, and the final covariance matrix is obtained by averaging over all realizations. Averaging across realizations significantly reduces the noise in the $w_p$ covariance matrix used for the cosmological analysis.

\begin{figure}
\includegraphics[width=1.0\columnwidth]{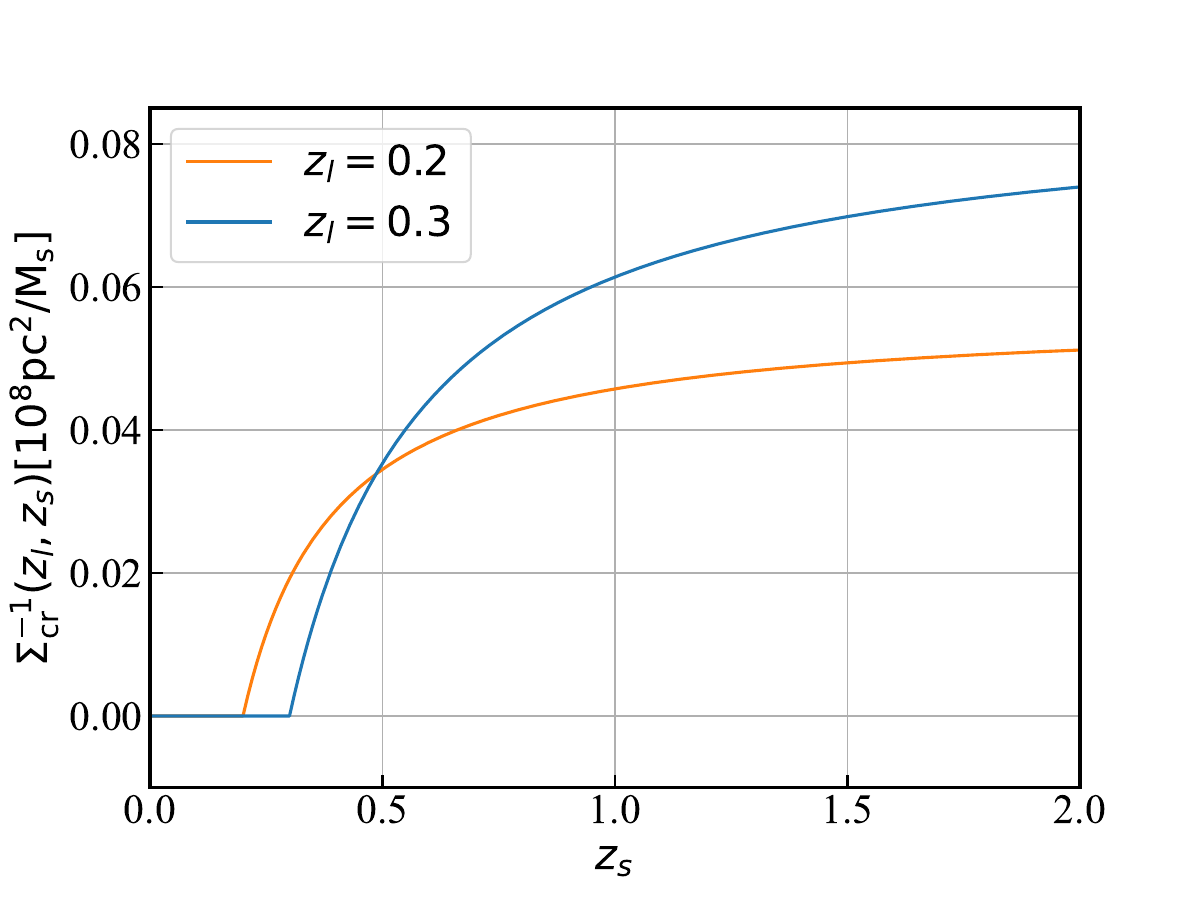}
\caption{\label{fig:sigma_cr_inv_demo} A demonstration of $\Sigma_{\text{cr}}(z_l, z_s)^{-1}$ with $z_l = 0.2$ and $z_l = 0.3$, which is proportional to $\gamma_t$, the quantity that governs the overall lensing signal-to-noise. This also shows the redshift dependency of the galaxy-galaxy lensing signal, which gives galaxy-galaxy lensing constraining power on the overall redshift distribution of the source sample, which we will see in Section~\ref{sec:results:0}.  }
\end{figure}

\subsection{\label{sec:measurement:ggl} Galaxy-Galaxy Lensing}

\begin{figure*}
\includegraphics[width=2.0\columnwidth]{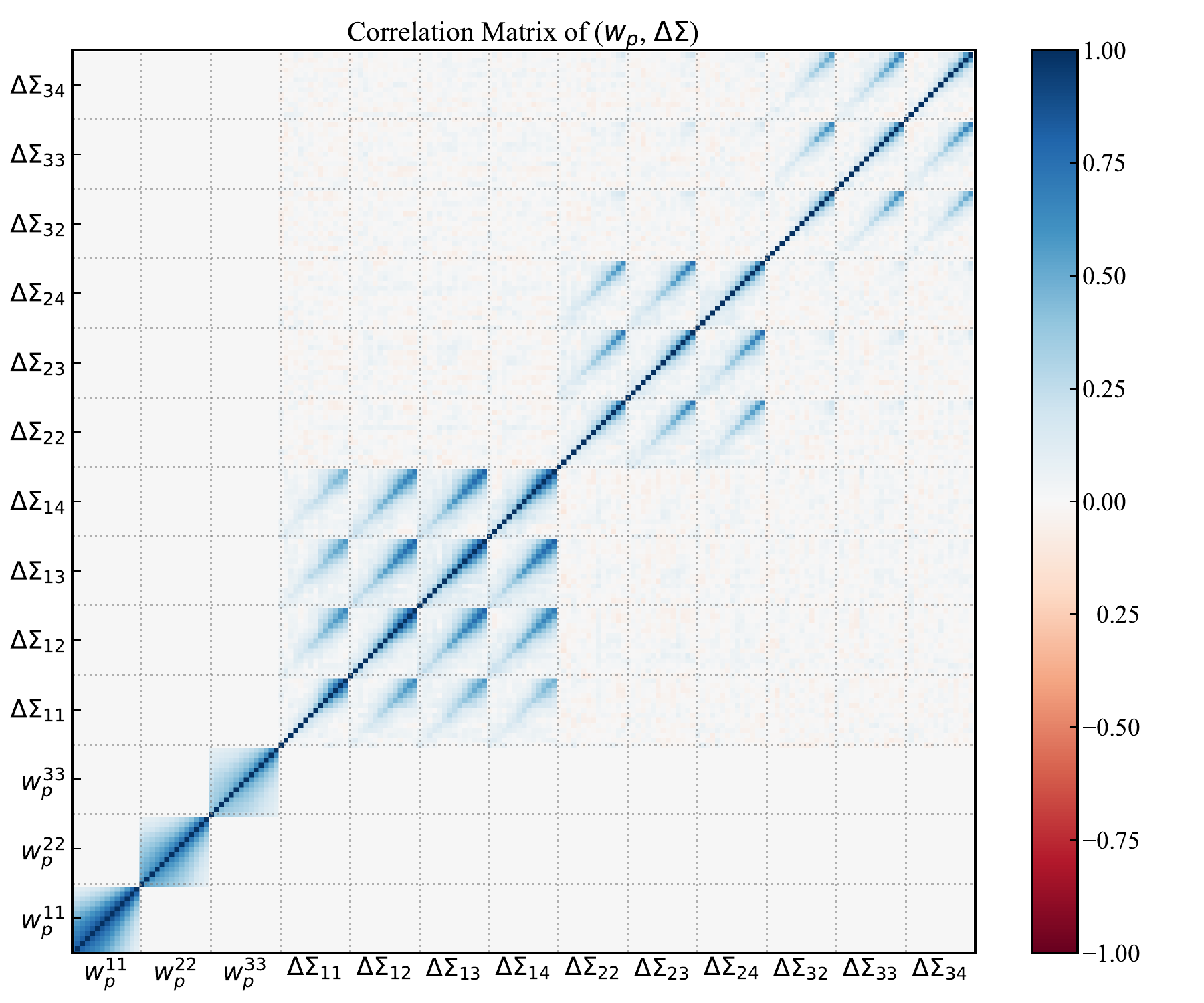}
\caption{\label{fig:corr_matrix}  The correlation matrix \textit{CORR} of the full data vector used in this work, estimated from the mock catalogs described in Section~\ref{sec:data:mock}. The matrix includes both the galaxy clustering measurements $w_p$ in three tomographic lens bins (LOWZ, CMASS1, CMASS2) and the galaxy-galaxy lensing measurements $\Delta \Sigma^{ij}$ across ten lens-source bin combinations. As shown, strong correlations exist within individual probes (e.g., among $w_p$ bins or $\Delta \Sigma$ bins) and off-diagonal bin pairs with the same lens bin. }
\end{figure*}

Predicted by General Relativity, mass distributions in the foreground cause deflections of background light, altering the apparent positions, fluxes, sizes, and shapes of background galaxies. This phenomenon is known as gravitational lensing. 
While weak gravitational lensing arises from all mass distributions along the line of sight, a useful signal for probing the large-scale structure is the lensing effect around foreground galaxies, referred to as galaxy-galaxy lensing.
% The most ubiquitous lensing effect in the universe occurs around foreground galaxies residing in halos—clumps of dark matter. To observe this effect, one measures the tangential shear around foreground galaxies, a measurement referred to as the galaxy-galaxy lensing signal.

To understand the observable of the galaxy-galaxy lensing, we first describe how weak lensing shear is estimated in the HSC Y3 shape catalog \cite{Li2022}. Weak lensing shear is estimated by the shape of the galaxies, which is a two-component complex number
\begin{equation}
\label{eq:shape_define}
e = e_1 + i e_2.
\end{equation}
$e_1$ defines the distortion along the x- and y-axes of the coadded images, while $e_2$ defines the distortion along the diagonals of the image. Taking a foreground lens galaxy as the center and considering a nearby source galaxy, one can describe the vector pointing from the lens to the source in polar coordinates $(R_p, \phi)$, where $R_p$ is the projected distance from the lens, determined by the angular diameter distance between the lens and the source and the redshift of the lens galaxy $z_l$, and $\phi$ is the polar angle of the vector. The tangential shear $\gamma_t$ \cite{bartelmann2001} can be described by
\begin{equation}
\label{eq:tangential_shear}
\gamma_t = \frac{1}{2\mathcal{R}}e_t = - \frac{1}{2\mathcal{R}}(e_1 \cos{(2\phi)} + e_2 \sin{(2\phi)}), 
\end{equation}
According to General Relativity, at a given projected radius $R_p$, the average tangential shear $\gamma_t (R_p)$ is given by the surface mass density profile $\Sigma(R_p)$ around the lens galaxy
\begin{equation}
\gamma_t (R_p) = \frac{\Sigma (<R_p) - \Sigma(R_p)}{\Sigma_{\text{cr}}(z_l, z_s)} = \frac{\Delta \Sigma (R_p)}{\Sigma_{\text{cr}}(z_l, z_s)}.
\end{equation}
Here $\Sigma (<R_p)$ denotes the average surface mass density within radius $R_p$, $\Delta \Sigma (R_p)$ denotes the excessive surface mass density, $\Sigma_{\text{cr}}(z_l, z_s)$ is the critical surface density
\begin{equation}
\label{eq:sigma_cr}
\Sigma_{\text{cr}}(z_l, z_s) = \frac{c^2}{4\pi G} \frac{D_A(z_s)}{ D_A(z_l) D_A(z_l, z_s) (1+z_l)^2},
\end{equation}
where $D_A(z_l)$, $D_A(z_s)$, and $D_A(z_l,z_s)$ denote the angular distance to the lens, to the source, and between the two respectively. 

Our estimator for the galaxy-galaxy lensing $\widehat{\Delta \Sigma} (R_p)$ \cite{sheldon2004, Mandelbaum2005} is defined as 
\begin{equation}
\label{eq:dsigma_def}
\Delta \Sigma (R_p) = \frac{1}{1+\hat{m}} \left(   \frac{\sum_{\text{ls}} w_{\text{ls}} e_{\text{t, ls}} \langle \Sigma_{\text{cr}}^{-1} \rangle_{\text{ls}}^{-1}}{ 2 \mathcal{R} \sum_{\text{ls} w_{\text{ls}}}} \right).
\end{equation}
Here we explain the quantities in Eq.~\ref{eq:dsigma_def}. 
\begin{enumerate}
    \item $\langle \Sigma_{\text{cr}}^{-1} \rangle_{\text{ls}}$: For each lens-source pair, the critical surface density is an integration over  Eq.~\ref{eq:sigma_cr} and the probability density function of the source photometric redshift $p(z_s)$.
    \begin{equation}
    \label{eq:sigma_inv_ls}
    \langle \Sigma_{\text{cr}}^{-1} \rangle_{\text{ls}} = \frac{4 \pi G (1+z_l)^2}{c^2} \bigintssss_{z_l}^{\infty} \frac{D_A(z_l) D_A{z_l, z_s}}{D_A(z_s)} p(z_s) \mathrm{d}z_s.
    \end{equation}
    For the $\Delta \Sigma (R_p)$ measurement, we use the $p(z_s)$ estimated by dNNz. With a fixed $z_l$, $\Sigma_{\text{cr}}(z_l, z_s)^{-1}$ is zero when $z_s < z_l$, and increases with $z_s > z_l$. This is demonstrated in Fig.~\ref{fig:sigma_cr_inv_demo}. Because the tangential shear $\gamma_t$ is proportional to $\Sigma_{\text{cr}}(z_l, z_s)^{-1}$ for a given excess surface mass density, $\Sigma_{\text{cr}}(z_l, z_s)^{-1}$ directly determines the amplitude and signal-to-noise ratio of the galaxy-galaxy lensing signal. Since $\Sigma_{\text{cr}}(z_l, z_s)^{-1}$ depends on the source redshift, it is sensitive to the source redshift distribution $n(z_s)$. The fiducial photo-$z$ method adopted in this work is \textsc{DNNz}, maintaining consistency with the HSC Y3 cosmic shear analyses \cite{Li2023, Dalal2023}. Additionally, we perform internal consistency tests using source redshifts estimated by \textsc{Mizuki} and \textsc{DEmPz} to ensure that the choice of photo-$z$ estimator does not significantly affect the cosmological results.
    \item $w_{\text{ls}}$: we apply optimum weighting for each lens-source pair by $w_{\text{ls}} = w_l w_s \langle \Sigma_{\text{cr}}^{-1} \rangle_{\text{ls}}^2$. As shown in Fig.~\ref{fig:sigma_cr_inv_demo}, the source galaxy further behind the lens will have a stronger tangential shear $\gamma_t$ as well as a higher lensing weight. Here $w_l$ is a weight on the lens galaxy to account for the inverse correlation between the number density of the galaxies and that of stars, and that of seeing, provided by SDSS DR11. $w_s$ is the weight of the source galaxies and is provided by the HSC Y3 shape catalog \cite{Li2022},
    \begin{equation}
    \label{eq:w_s} w_s = \frac{1}{\sigma_{e;s}^2 + e^2_{\text{RMS};s}}, 
    \end{equation}
    where $\sigma_{e;s}$ is the per-component $1\sigma$ uncertainty of the shape estimation due to the Poisson noise of the photon, $e_{\text{RMS};s}$ is the per-component root-mean-square (RMS) of the galaxy intrinsic ellipticity. Both quantities are estimated for each galaxy using the image simulation. In the shape noise-dominated limit, the weights minimize the variance of the galaxy-galaxy lensing estimator, by up-weighting sources with low measurement uncertainties \cite{Hirata2003}, and up-weighting lens-source pairs with greater lensing efficiency. 
    \item The shear responsivity $\mathcal{R}$ is a quantification of how the ellipticity estimator $e$ react to weak lensing shear $\gamma$\cite{Jarvis2003}. It is estimated by
    \begin{equation}
    \label{eq:responsivity} \mathcal{R} = 1 - \frac{\sum_{\text{ls}} w_{\rm ls} e^2_{\text{RMS}}}{ \sum_{\text{ls}} w_{\rm ls}}.
    \end{equation}
    \item The average multiplicative bias $\hat{m}$ is estimated by
    \begin{equation}
    \label{eq:multiplicative_bias} \hat{m} = \frac{\sum_{\text{ls}} w_{\rm ls} m_s}{ \sum_{\text{ls}} w_{\rm ls}},
    \end{equation}
    where $m_s$ is the per-galaxy source multiplicative bias calculated in Eq.~\ref{eq:mbias_blinding}.
\end{enumerate}

Additionally, we apply ensemble additive and multiplicative selection bias corrections following Li \textit{et al.} \cite{Li2022}. We adopt the $\hat{a}_{\rm sel}$ and $\hat{m}_{\rm sel}$ values from the HSC Y3 cosmic shear analysis and apply them to the $\Delta \Sigma$ measurement according to
% \begin{equation}
% \label{eq:dsigma_selection_bias} \Delta \Sigma \rightarrow\frac{1}{1+\hat{m}_{\rm sel} (\Delta \Sigma - \hat{a} \Delta\Sigma^{\rm PSF})}.
% \end{equation}
\begin{equation}
\label{eq:dsigma_selection_bias} \widehat{\Delta \Sigma} \rightarrow\frac{1}{1+\hat{m}_{\rm sel}}(\widehat{\Delta \Sigma} - \hat{a}_{\rm sel} \widehat{\Delta \Sigma}^{\rm PSF}).
\end{equation}

Here $\Delta \Sigma^{\rm PSF}$ is expressed as 
\begin{equation}
\label{eq:delta_sigma_psf} \widehat{\Delta \Sigma}^{\rm PSF} = \frac{\sum_{\rm ls} w_{\rm ls} e_{\rm t, ls}^{\rm PSF}}{\sum_{\rm ls} w_{\rm ls}},
\end{equation}
where $e_{\rm t, ls}^{\rm PSF}$ is the tangential component of the PSF model shape at the galaxy's location.

We repeat the same measurement procedure for the weak lensing signal, but around random points instead of lens galaxies. The random catalog contains 40 times more points than the number of lens galaxies. The random points occupy the same footprint as the lens galaxies and are generated to follow the same redshift distribution. As shown in Fig.~\ref{fig:dsigma_cross}, the measurement around the randoms, denoted as $\Delta \Sigma_{\rm rand}$, captures the systematic cross term. Since the randoms capture this signal up to $70h^{-1}$Mpc for the cross term, we accordingly correct the tangential estimator by
\begin{equation}
\label{eq:subtract_random} \widehat{\Delta \Sigma}(R_p) \rightarrow \widehat{\Delta \Sigma} - \widehat{\Delta \Sigma}_{\rm rand}(R_p).
\end{equation}

Finally, since we are using source redshift bins that overlap with the lens redshift bins, some source galaxies that are actually within the lens redshift range are inevitably misidentified as background sources and assigned nonzero weights by $\langle \Sigma_{\text{cr}}^{-1} \rangle_{\text{ls}}^2$. 
These galaxies do not experience lensing by the lens halo but instead dilute the lensing signal\cite{sheldon2009, Miyatake2015}. We detect and correct for this contamination using the boost factor, which quantifies the excess source galaxies around lenses relative to random points. We note that the boost factor correction specifically addresses associated galaxies and does not correct for other sources of dilution, such as foreground galaxies due to photometric redshift errors. 

\begin{figure*}
\includegraphics[width=2.0\columnwidth]{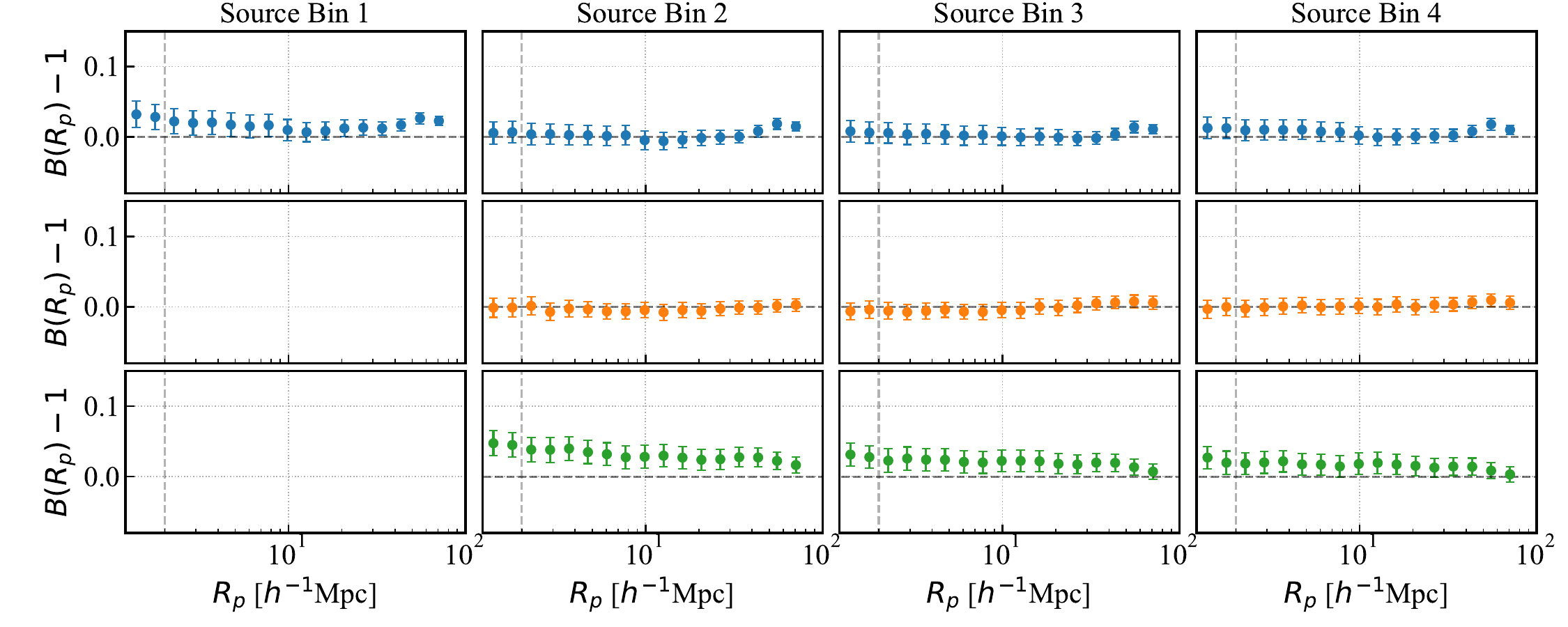}
\caption{\label{fig:boost_factor} The boost factor $B(R_p)-1$ of the lens-source bin pair. The three rows from top to low panels correspond to LOWZ, CMASS1, and CMASS2 as lens bins, and the columns correspond to the HSC 1-4 bin. $B(R_p) - 1$ shows the ratio of the source galaxies that are physically associated with the lenses.  %\rachel{Consider plotting $B(R_p)-1$, to focus on the physical interpretation?} \tianqing{Done.}
}
\end{figure*}

The boost factor is defined as
\begin{equation}
\label{eq:boost_factor}   B(R_p) = \frac{ \sum_{\rm ls} w_{wl} / \sum_l w_l}{ \sum_{rs} w_{rs} / \sum_r w_r}
\end{equation}
Here, $w_r$ is the weight assigned to random points and is set to 1 for all points. The random–source pairing weight is computed as $w_{\rm rs} = w_r w_s \langle \Sigma_{\text{cr}}^{-1} \rangle_{\rm rs}^2$. For an ideal source catalog that is well-separated with the lens redshift, $B(R_p) = 1$. For a source sample that has contaminated galaxies from the lens redshift, $B(R_p)>1$ because those contaminating galaxies are clustering with the lens sample.  The quantity $B(R_p) - 1$ represents the fraction of source galaxies that are mistakenly treated as background sources but are in fact physically associated with the lens galaxies. We correct the galaxy-galaxy lensing signal by applying the boost factor as
\begin{equation}
\label{eq:correct_boost} \widehat{\Delta \Sigma}(R_p) \rightarrow  B(R_p) \widehat{\Delta \Sigma} (R_p).
\end{equation}
In Fig.~\ref{fig:boost_factor}, we show $B(R_p)$ for each lens-source bin in our galaxy-galaxy lensing measurement. We can see that the small scales for LOWZ-HSC1, CMASS2-HSC2,3, and 4 have $B(R_p)$ consistently larger than 1. This means some portion of those source samples are un-lensed contamination, and we adjust the data vector $\widehat{\Delta \Sigma}$ by Eq.~\ref{eq:correct_boost}. 
% \sunao{Is this applied in mock meausrmeent as well? Does the covairance include the scatter of the boost factor? My previous analysis did not do this, so I am wondering if it has any effects.}\tianqing{I think your previous analysis don't need to do it because boost factor is small for the single source bin.}

We note that the boost factor correction assume that the tangential shear of source galaxies physically associated to the lens is zero, which is broken by the existence of intrinsic alignment\cite{Bridle2007, Heymans2013}.

The lower panel of Fig.~\ref{fig:measurement} shows the $\Delta \Sigma$ measurements. We measure $\Delta \Sigma^{ij}(R_p)$, where $i$ denotes the lens bin index—1 = LOWZ, 2 = CMASS1, and 3 = CMASS2—and $j$ denotes the HSC Y3 source bin index. Each $\Delta \Sigma^{ij}(R_p)$ measurement is performed in 30 radial bins, equally spaced in $\log(R_p)$ between $0.05 < R_p < 80\ h^{-1}\mathrm{Mpc}$. Throughout the paper, the galaxy-galaxy lensing data vector is arranged in the following order: $\Delta \Sigma^{11}$, $\Delta \Sigma^{12}$, $\Delta \Sigma^{13}$, $\Delta \Sigma^{14}$, $\Delta \Sigma^{22}$, $\Delta \Sigma^{23}$, $\Delta \Sigma^{24}$, $\Delta \Sigma^{32}$, $\Delta \Sigma^{33}$, $\Delta \Sigma^{34}$. We drop $\Delta \Sigma^{21}$ and $\Delta \Sigma^{31}$ due to the excessive redshift overlap between the lens and source bins.

The error bars shown in Fig.~\ref{fig:measurement} for the galaxy-galaxy lensing signal are computed from the diagonal elements of the covariance matrix. The covariance matrix for $\Delta \Sigma(R_p)$ is estimated using the measurements from 1404 mock catalogs of HSC Y3 source galaxies, as described in Section~\ref{sec:data:mock}. The covariance is computed as 
\begin{equation} 
\label{eq:covariance_dsds} 
\mathcal{C}^{ij, uv}(R_1, R_2) = \langle \widehat{\Delta \Sigma}^{ij}(R_1) \widehat{\Delta \Sigma}^{uv}(R_2) \rangle, 
\end{equation} 
where $i$ and $u$ are the lens bin indices, and $j$ and $v$ are the source bin indices.
We assume no correlation between the $w_p$ and $\Delta \Sigma$ data vectors because the overlap between the HSC Y3 and SDSS footprints is negligible compared to the full SDSS area. The full correlation matrix of $w_p$ and $\Delta \Sigma$ is shown in Fig.~\ref{fig:corr_matrix}.

\subsection{\label{sec:measurement:null} Null Testing}

The clustering signal measurement in this work is identical to that in More \textit{et al.} \cite{More2023}. Therefore, we do not repeat the null tests or systematics tests for the $w_p$ data vector and refer the reader to More \textit{ et al.} \cite{More2023} for details.

For the galaxy-galaxy lensing signal, we repeat a subset of the null tests performed in the previous HSC Y3 analysis to ensure that our galaxy-galaxy lensing data vector, constructed using the tomographic source sample, is not significantly contaminated by systematic effects.

% The HSC Y3 source galaxies are divided into tomographic bins based on their point estimates of photometric redshift and therefore carry significant uncertainties. These uncertainties lead to long tails in the redshift distribution, as seen in Fig.~\ref{fig:redshift_distro}. As a result, some source galaxies are incorrectly placed behind the lens galaxies but are actually at the same redshift as the lenses. We can detect the fraction of such galaxies using the "boost factor," which is the ratio of source galaxy counts around lenses to that around random points \cite{sheldon2009, Miyatake2015}. The boost factor is defined as
% \begin{equation}
% \label{boost_factor_def} B(R) = \frac{\sum_{ls} w_{ls} / \sum_l w_l}{\sum_{rs} w_{rs} / \sum_r w_r},
% \end{equation}
% where ``r'' stands for random catalogs, and $w_r$ its weight. For an ideal source catalog that is well-separated with the lens redshift, $B(R) = 1$. For a source sample that has contaminated galaxies from the lens redshift, $B(R)>1$ because those contaminating galaxies are clustering with the lens sample.  $B(R)-1$ is the ratio of galaxies misidentified as background galaxies with a particular lens ensemble.  

\begin{figure*}
\includegraphics[width=2.0\columnwidth]{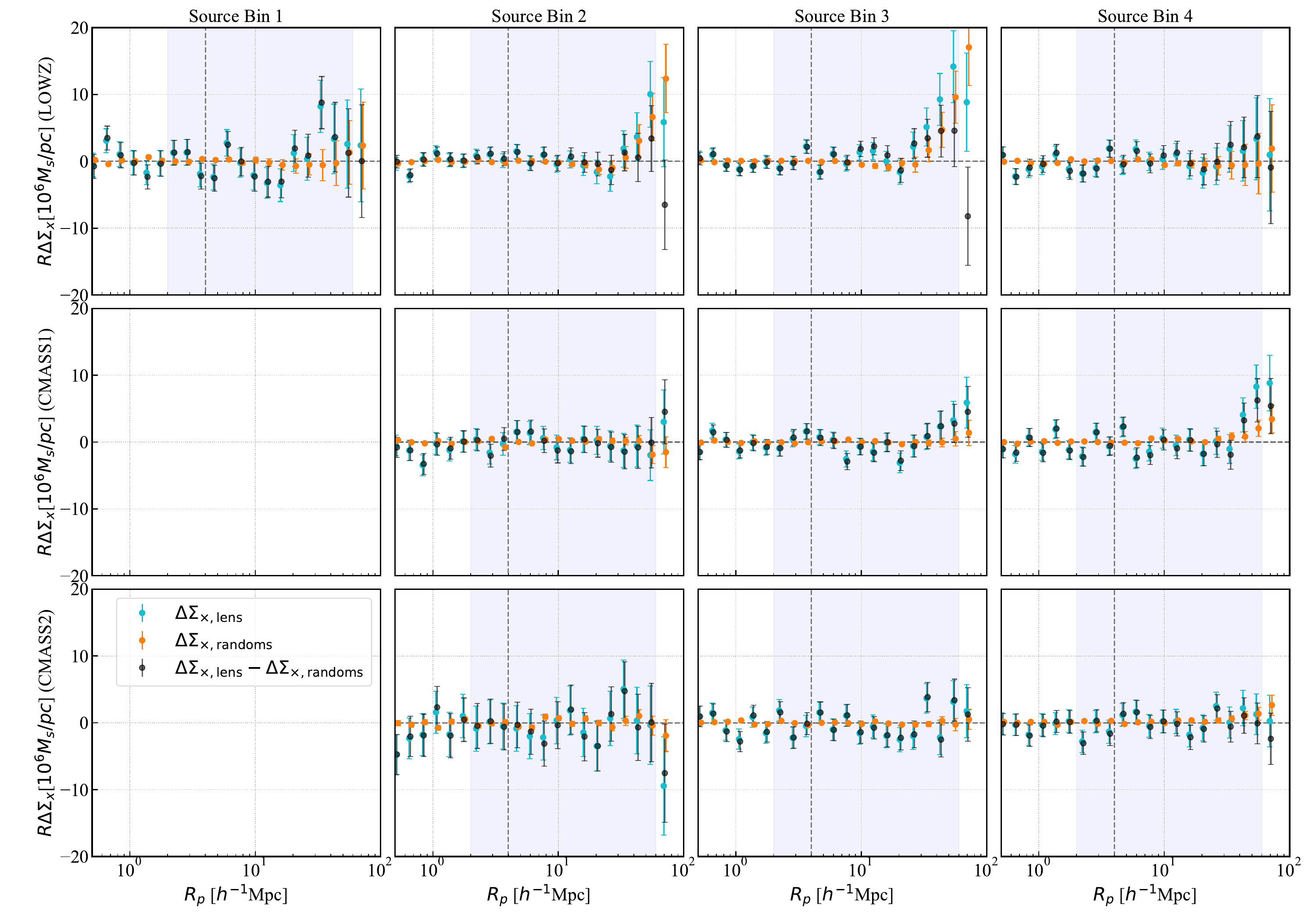}
\caption{\label{fig:dsigma_cross} The cross component of the weak lensing signal. The blue points are the cross-component around the lens catalog, $\Delta \Sigma_{x, {\rm lens}}$, and the orange points are the cross-component around the random catalog, $\Delta \Sigma_{x, {\rm rand}}$, the black points are the cross-component of the estimator  $\widehat{\Delta \Sigma}_x$, which is the difference between the blue and the orange. The angular range where $\widehat{\Delta \Sigma}_x$ are consistent with zero is deemed to be free of significant contamination. The blue shaded regions indicate the scales used in this work for cosmological inference. The dashed lines in the $\Delta \Sigma$ panels denote the scales of $R_0$ for the point mass correction term.  %\rachel{It is difficult to distinguish between blue and black; consider another color combination?  Also, consider putting a legend into one of the empty panels?  Need to explain the shaded regions.} \tianqing{slightly shifted the points and change the tint of the blue and black points, add legend, explained the shaded regions. }
}
\end{figure*}

\begin{figure*}
\includegraphics[width=2.0\columnwidth]{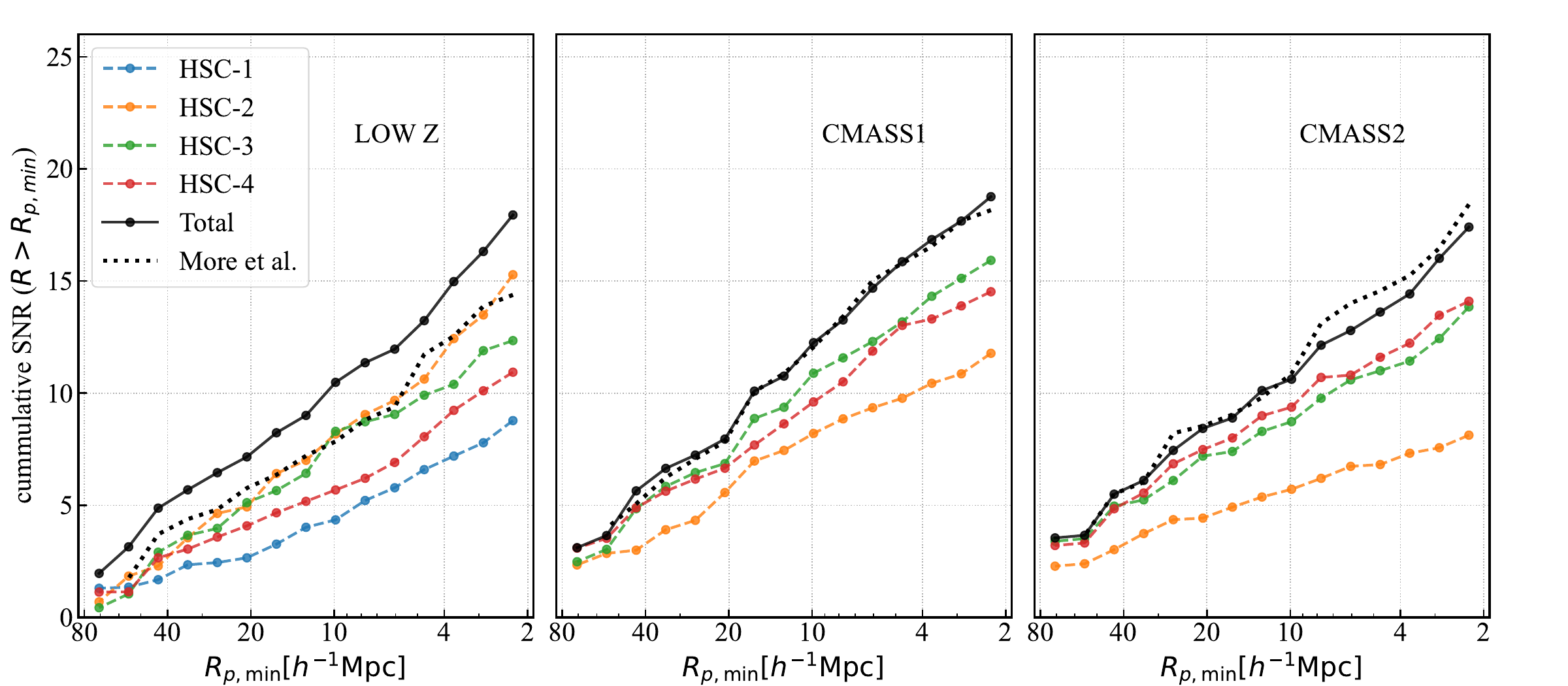}
\caption{\label{fig:signal_to_noise} The cumulative signal-to-noise ratio (SNR) of the galaxy-galaxy lensing using tomographic source samples. The colored lines show the SNR between $R_{p,\rm min} < R_p < 80 h^{-1}$Mpc for individual len-source bin pairs. The black lines show the overall SNR of a single lens bin with all its available source bins. The black dotted line shows the SNR measure using one high redshift source bin in \cite{More2023}. In the radius range of $12-80 h^{-1}$Mpc, 
we gain 25\% in LOWZ in signal-to-noise compared to the single bin, while the single bin remains as effective for the CMASS1 and CMASS2 measurements.  
% \rachel{This is by comparing the bin 4 curves versus total?} \tianqing{Compared to the one redshift bin used in HSC Y3} \rachel{So it's not something we can actually see from this figure?  That is a bit confusing to have as a major takeaway.  Personally, I would recommend putting some kind of symbol on the plot (like a single large point) for the HSC Y3 analysis given its range of scales used and single redshift bin.  Then you can make it clear that is where the numbers are coming from.}\tianqing{I overlaied the single bin SNR, turns out it is not really that much of a gain except for LOWZ.} %\rachel{What are the different line styles meant to show?  It's not clear to me and the caption doesn't mention them.  Also, please consider switching to a colorblind-friendly color palette.  The caption and y axis label use different notation (S/N versus SNR) - should homogenize this.} \tianqing{Remove the different line styles, change to colorblind-friendly palette, and change the y-label here. }
}
\end{figure*}

\begin{table*}[ht]
\centering
\caption{Summary of the $\chi^2$/dof and $p$-values for the galaxy-galaxy lensing cross-component null tests, evaluated across the scale of [2,70]$h^{-1}$Mpc. Results are shown for individual lens bins and the total data vector, evaluated with three different photometric redshift methods.}
\label{tab:dsigma_cross_chi2}
\begin{tabular}{lccc}
\hline\hline
Lens Bin & DNNz & Mizuki & DEmPz \\
\hline
LOWZ        & $\chi^2/$dof = 61.3 / 56 (p = 0.28) & $\chi^2/$dof = 59.9 / 56 (p = 0.33) & $\chi^2/$dof = 58.0 / 56 (p = 0.41) \\
CMASS1      & $\chi^2/$dof = 40.4 / 42 (p = 0.53) & $\chi^2/$dof = 39.2 / 42 (p = 0.59) & $\chi^2/$dof = 38.5 / 42 (p = 0.62) \\
CMASS2      & $\chi^2/$dof = 31.9 / 42 (p = 0.87) & $\chi^2/$dof = 30.3 / 42 (p = 0.91) & $\chi^2/$dof = 29.1 / 42 (p = 0.94) \\
\hline
Total       & $\chi^2/$dof = 148.7 / 140 (p = 0.29) & $\chi^2/$dof = 145.7 / 140 (p = 0.35) & $\chi^2/$dof = 139.5 / 140 (p = 0.49) \\
\hline
\end{tabular}
\end{table*}

Another null signal we inspect for systematics is the cross component of the weak lensing signal, $\Delta \Sigma_x$, defined similarly to Eq.~\ref{eq:dsigma_def} but replacing $e_t$ with $e_x$. The cross component of the estimator is given by $\widehat{\Delta \Sigma}_x = \Delta \Sigma_{x, {\rm lens}} - \Delta \Sigma_{x, {\rm rand}}$, and all three terms measured using \textsc{DNNz} are shown in Fig.~\ref{fig:dsigma_cross}. The blue data points correspond to the cross component of the estimator, which is the difference between the cross components measured around the lenses and around the random catalog.
The cross components of the estimator are expected to be consistent with zero across the angular scales used in our analysis. We observe a significant upturn at large scales for the LOWZ and CMASS1 lens–source tomographic bin pairs around the lenses; however, the cross components measured around the random catalog trace the same upturn, and they cancel out in the final estimator. We remove the last data points in all tomographic bin pairs, as including them would significantly worsen the $\chi^2$ of the fit, and drop LOWZ-HSC3's $p$-value to $0.002$. This large-scale upturn does not appear in the CMASS2 tomographic bin pairs, likely because the CMASS2 samples are sensitive to smaller radius scales \cite{More2023}. 

We summarize the results of the systematics tests for the galaxy-galaxy lensing signal by computing the chi-square per degree of freedom ($\chi^2$/dof) and the corresponding $p$-values for the null signal in Table~\ref{tab:dsigma_cross_chi2}. We find that all $p$-values for the null tests, whether evaluated for individual lens bins or for the full data vector, are greater than 0.05 for all three photometric redshift methods. This suggests that our data vector is unlikely to be significantly contaminated by systematics within the range of scales used for the cosmological analysis.

Other systematics tests that depend on the choice of lens catalog are not repeated in this work, since we use the same lens catalog as in More \textit{et al.} \cite{More2023}. These tests include, for example, examining the slope of the lensing signal as a function of redshift within the lens bins, and studying the variation of the galaxy-galaxy lensing signal with and without applying lens weights. We refer readers to More \textit{et al.} \cite{More2023} for further details on these systematics tests.

In Fig.~\ref{fig:signal_to_noise}, we show the cumulative signal-to-noise ratio (SNR) of the tomographic galaxy-galaxy lensing data vector, evaluated between a minimum radius $R_{p,\rm min}$ and a maximum radius of $70\ h^{-1}\mathrm{Mpc}$. The total SNR for each lens bin is shown with a black dotted line, while the contributions from individual lens–source bin pairs are shown with colored lines. The two highest redshift HSC source bins contribute the most SNR to the CMASS1 and CMASS2 samples, while the second and third source bins dominate the SNR for LOWZ. 
Compared to using a source sample restricted to $z > 0.75$, as in the previous HSC Y3 3×2pt analysis \cite{More2023, Sugiyama2023, Miyatake2023}, the tomographic galaxy-galaxy lensing only have significant gain in SNR within the same radius range for LOWZ by about 25\%, since more source galaxies are excluded by a high-$z$ cut. 
% This demonstrates that including the lower-redshift source bins increases the overall SNR.
% Compared to using a source sample restricted to $z > 0.75$, as in the previous HSC Y3 3×2pt analysis \cite{More2023, Sugiyama2023, Miyatake2023}, we achieve SNR gains of 47\%, 14\%, and 24\% for the LOWZ, CMASS1, and CMASS2 bins, respectively. The improvement is especially significant for LOWZ, since more source galaxies are excluded by a high-$z$ cut in that case.

\section{\label{sec:model:0} Modeling and Analysis Choices}

In this section, we describe the models and analysis methods used in this work. This includes the theoretical modeling of galaxy clustering and galaxy-galaxy lensing, treatments for systematic errors, the framework for Bayesian inference, and the validation and internal consistency tests performed to ensure the robustness of the results.

\subsection{\label{sec:model:theory} Thoretical Model}

\subsubsection{\label{sec:model:theory_wp} Projected correlation function $w_p(R_p)$}

We model the galaxy clustering of the SDSS BOSS lens galaxies using the “minimal bias” model \cite{Sugiyama2022, Sugiyama2023}. In this model, the galaxy number density is related to the underlying matter density through a linear galaxy bias, such that the galaxy overdensity is given by $\delta_g = b_l(z_l) \delta_m$, where $b_l(z_l)$ is the linear galaxy bias at redshift $z_l$, and $\delta_m$ is the matter overdensity. The minimal bias model has been validated for modeling the galaxy clustering signal in the HSC Y3 analysis at scales larger than $8\ h^{-1}\mathrm{Mpc}$ \cite{Sugiyama2022}.

In this formalism, we first compute the 3D galaxy correlation function $\xi_{gg}$ as 
\begin{equation} 
\xi_{gg}(r; z_l) = b_l(z_l)^2 \int_0^{\infty} \frac{k^2 \,\mathrm{d}k}{2\pi^2} P_{\rm mm}^{\rm NL}(k; z_l)\, j_0(kr), 
\end{equation} 
where $P_{\rm mm}^{\rm NL}(k; z_l)$ is the nonlinear matter power spectrum at wavenumber $k$ and redshift $z_l$, and $j_0(x)$ is the zeroth-order spherical Bessel function. Throughout this work, we use \textsc{CAMB} \cite{Lewis2000} to compute the linear matter power spectrum and apply \textsc{HaloFit} \cite{Takahashi2012} for its nonlinear correction.
We do not include baryonic feedback effects in this analysis, as they primarily impact the cosmic shear signal, which is not used in this work.
We do not include baryonic feedback effects in this analysis. While baryonic processes can alter the matter distribution on small scales and impact galaxy-galaxy lensing signals \cite{huang2019}, their effects are suppressed at the relatively large scales ($R_p>2h^{-1}$Mpc) used in this work. Our choice of scale cuts mitigates the impact of baryonic uncertainties on the lensing signal, and are confirmed by validation tests shown in Section~\ref{sec:validation:validation}.

The projected auto-correlation function of the SDSS galaxies, $w_p(R_p; z_l)$, is computed by integrating the 3D galaxy correlation function $\xi_{gg}(r; z_l)$ along the line of sight: 
\begin{equation} 
w_p(R_p; z_l) = 2 f^{\rm RSD}_{\rm corr}(R_p; z_l) \int_0^{\Pi_{\rm max}} \mathrm{d}\Pi\, \xi_{gg} \left(\sqrt{R_p^2 + \Pi^2}; z_l\right). 
\end{equation} 
We fix the maximum line-of-sight integration limit to $\Pi_{\rm max} = 100\ h^{-1}\mathrm{Mpc}$ throughout this work, following \cite{vandenbosch2013}. The 3D separation $r$ is expressed as a combination of the projected transverse separation $R_p$ and the line-of-sight separation $\Pi$. The factor $f^{\rm RSD}_{\rm corr}(R_p; z_l)$ accounts for the correction due to the Kaiser redshift-space distortion (RSD) effect along the line of sight [See Eq.(48) in \cite{vandenbosch2013} for definition]. This factor is on the order of 1 to 1.2 in the $R_p$ range of this work. 

Following \cite{Sugiyama2023}, we ignore the redshift evolution of the clustering signal. The auto-correlation function $w_p(R_p; z_l)$ for the LOWZ, CMASS1, and CMASS2 lens galaxies is evaluated at fixed redshifts of $z_l = 0.26$, $0.51$, and $0.63$, respectively. The redshift evolution of $w_p$ assuming a redshift-independent galaxy bias $b_l(z_l)$ contributes at most 4\% of the statistical uncertainty in $w_p$ \cite{More2023}, and can therefore be safely neglected.

\subsubsection{\label{sec:model:theory_ggl} Galaxy-galaxy lensing $\Delta \Sigma (R_p)$}

In this work, galaxy-galaxy lensing is measured via the excess surface mass density at projected radius $R_p$, denoted as $\Delta \Sigma(R_p)$. This quantity is derived from the directly observed tangential shear $\gamma_t$, and the redshifts of the lens and source galaxies, $z_l$ and $z_s$, as described in Section~\ref{sec:measurement:ggl}. The theoretical model for $\Delta \Sigma(R_p)$ includes three components: \begin{equation} \label{eq:ggl_theory_all} \Delta \Sigma(R_p) = \Delta \Sigma_{gG}(R_p) + \Delta \Sigma_{\rm PM}(R_p) + \Delta \Sigma_{\rm mag}(R_p). \end{equation} The first term, $\Delta \Sigma_{gG}(R_p)$, represents the excess surface density predicted using the nonlinear matter power spectrum $P_{\rm mm}^{\rm NL}(k; z_l)$ under the minimal bias model: 
\begin{equation} 
\label{eq:ggl_first_term} 
\Delta \Sigma_{gG}(R_p) = b_l(z_l)\, \bar{\rho}_{m,0} \int_0^\infty \frac{k\, \mathrm{d}k}{2\pi} P{\rm mm}^{\rm NL}(k; z_l), J_2(kR_p), 
\end{equation} 
where $\bar{\rho}_{m,0}$ is the mean matter density at $z=0$, and $J_2(x)$ is the second-order Bessel function.

\begin{figure}
\includegraphics[width=0.95\columnwidth]{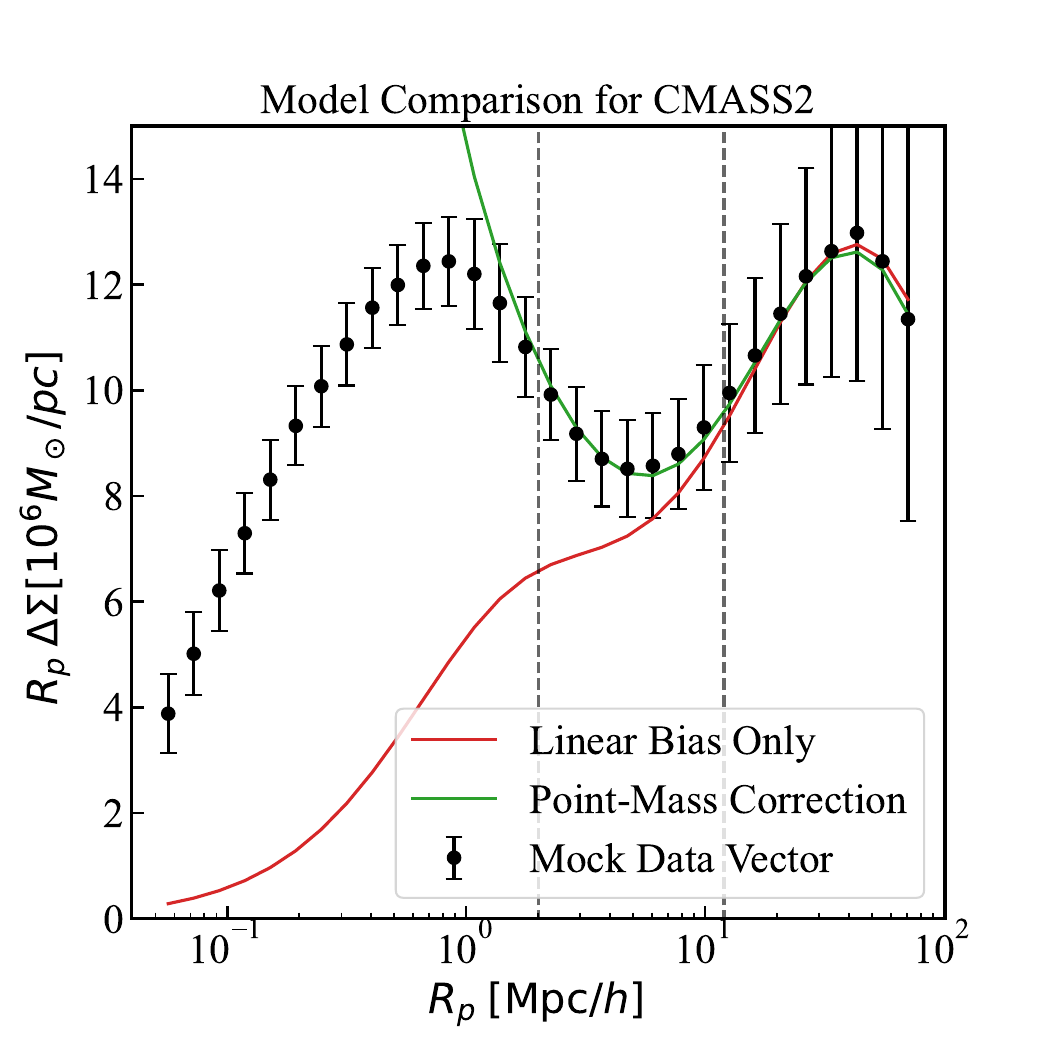}
\caption{\label{fig:model_comparison} This figures compares the best-fitting ``linear bias only'' model versus the ``point-mass correction'' model on the mock measurement of $R_p \Delta \Sigma$ of CMASS1. The ``linear bias only'' can only accurately describe the physical scale for  $R_p > 12 h^{-1}$Mpc (right dashed line), while the point-mass correction model can describe the data down to scales around $2 h^{-1}$Mpc. The error bars reflect the uncertainties of the real data vector, so it is important to model the noiseless mock within $1\sigma$ of the errorbar. 
% \rachel{y axis needs units.  Consider putting errorbars for the real data to give a sense of how close theory is with respect to them?} \tianqing{Done and done.} \rachel{Thanks!  I think you need to explain in the caption that the errorbars are from the data and do not reflect some uncertainty in the mocks based on cosmic variance (or something like that).  This is important to contextualize the importance of being well within $1\sigma$.}\tianqing{Done.}
}
\end{figure}

%\rachel{What's below needs references to Baldauf et al (2010) and Mandelbaum et al (2013) for definition and use of the ADSD.  Should also refer to and comment on the recent Prat et al paper comparing the ADSD approach, point mass model, and other modeling approaches for small-scale GGL.}  
%\rachel{Besides references, I think more explanation would be helpful here.  Specifically: what issue are you trying to solve / why is this more important for GGL than clustering?  What precisely are the assumptions of the method?  For example, some people imagine we are assuming the mass inside $R_0$ is distributed like a point mass, but we are not actually doing so.}\tianqing{Done}
%\rachel{Have you considered using a mock catalog signal to illustrate what the point mass term is doing?  For example, model the mock catalog signal with just the term from NL power spectrum with linear bias, and then with both terms: what happens to how well you can fit the signal on the scales you use?  What happens to the inferred cosmological parameters?  This could go in the model validation section later on.}\tianqing{Added figure 8 for this purpose. }
The second term is a correction to account for small-scale mis-modeling in the \textsc{HaloFit} nonlinear power spectrum. Since \textsc{HaloFit} does not include the 1-halo term in its formalism, its prediction for $\Delta \Sigma_{gG}(R_p)$ falls off too rapidly at small $R_p$, leading to deviations from the observed measurements. One approach to reduce sensitivity to poorly modeled small-scale contributions is the Annular Differential Surface Density (ADSD), typically denoted as $\Upsilon(R_p)$, introduced in \cite{Baldauf2010} and later applied to data in \cite{Mandelbaum2013}. It is defined as
\begin{equation}
\label{eq:upsilon}
\Upsilon(R_p) = \Delta \Sigma_{gG}(R_p) - \Delta \Sigma_{gG}(R_0) \left(\frac{R_0}{R_p}\right)^2.
\end{equation}
where $R_0$ is the inner radius of the annulus. The second term removes the contribution from the mass distribution inside $R_0$, making $\Upsilon(R_p)$ insensitive to the small-scale matter profile.
In this work, we reverse Eq.~\ref{eq:upsilon} to construct a corrected version of $\Delta \Sigma_{gG}(R_p)$ that includes a free parameter representing the missing 1-halo contribution:
\begin{align}
    \Delta \Sigma'(R_p)  &= \Upsilon(R_p) + \Delta \Sigma_{\rm 1-halo}(R_0) (\frac{R_0}{R_p})^2\\
                        &= \Delta \Sigma_{gG}(R_p) + \left[\Delta \Sigma_{\rm 1-halo}(R_0) - \Delta \Sigma_{gG}(R_0)\right] \left(\frac{R_0}{R_p}\right)^2. 
\end{align}
Here, $\Delta \Sigma_{\rm 1-halo}(R_0)$ is a free parameter capturing the excess surface density at $R_0$ due to the missing 1-halo term, and $\Delta \Sigma_{gG}(R_p)$ is the \textsc{HaloFit} prediction. We fix $R_0 = 4\ h^{-1}\mathrm{Mpc}$ throughout the analysis so that there are three data points below $R_0$ for a proper $\Delta \Sigma_{\rm 1-halo}(R_0)$ fitting. The resulting point-mass correction term in Eq.~\ref{eq:ggl_theory_all} is given by:
\begin{equation}
\label{eq:ggl_second_term}
\Delta \Sigma_{\rm PM}(R_p) = \left[\Delta \Sigma_{\rm 1-halo}(R_0) - \Delta \Sigma_{gG}(R_0)\right] \left(\frac{R_0}{R_p}\right)^2.
\end{equation}
It is worth noting that a similar strategy to marginalize over small-scale modeling uncertainty was used in the DES Y3 galaxy-galaxy lensing analysis \cite{Prat2022}, although a slightly different treatment was adopted due to their use of angular binning instead of physical radius binning.

In Fig.~\ref{fig:model_comparison}, we demonstrate the advantage of using the point-mass correction model in describing a mock data vector that includes physical features from an $N$-body simulation. With the addition of the point-mass correction term, the model successfully extends the minimum scale at which $\Delta \Sigma(R_p)$ can be accurately described -- from $12$ down to $2\ h^{-1}\mathrm{Mpc}$.
It is important to note that by including the point-mass correction term, we are not assuming that all mass within $R_0$ is concentrated at the center of the halo. Rather, we are asserting that $\Delta \Sigma(R_p)$ between $4$ and $12\ h^{-1}\mathrm{Mpc}$ can be reasonably approximated by this term, while remaining agnostic about the mass distribution within $2\ h^{-1}\mathrm{Mpc}$. 
% The data points between $2$ and $4\ h^{-1}\mathrm{Mpc}$ are effectively used to constrain the free parameter $\Delta \Sigma_{\rm 1-halo}(4\ h^{-1}\mathrm{Mpc})$.

The third term in Eq.~\ref{eq:ggl_theory_all} represents the magnification contribution, which describes an additional observed galaxy-galaxy lensing signal caused by additional number of lens galaxies located at over-density regions in the foreground because of the lensing magnification. It is given by
\begin{equation}
\label{eq:ggl_mag1}
\Delta \Sigma_{\rm mag} (R_p) = \int {\rm d}z_s n_s(z_s) \int {\rm d}z_l n_l(z_l) \widetilde{\Delta \Sigma}_{\rm mag}(R_p;z_l,z_s),
\end{equation}
where $n_s(z_s)$ and $n_l(z_l)$ are the redshift distributions of the source and lens galaxies, respectively. The integrand $\widetilde{\Delta \Sigma}_{\rm mag}(R_p; z_l, z_s)$ is expressed as
\begin{align}
\widetilde{\Delta \Sigma}_{\rm mag}(R_p;z_lmz_s) = &2(\alpha_{\rm mag} - 1) \\\nonumber & \int_0^\infty \frac{\ell {\rm d}\ell}{2\pi} \Sigma_{\rm cr}(z_l, z_s) C_{\kappa} J_2\left(\ell \frac{R_p}{\chi_l}\right).
\end{align}
$\Sigma_{\rm cr}(z_l, z_s)$ is the critical surface density from Eq.~\ref{eq:sigma_cr}. $\alpha_{\rm mag}$ is estimated by the slope of number counts at the luminosity cut of each lens bin. Following the previous study\cite{Sugiyama2023,Miyatake2023}, we assign a conservative $\sigma_\alpha = 0.5$ on each $\alpha_{\rm mag}$ parameter during the analysis process to account for uncertainty for the $\alpha_{\rm mag}$ estimation.   $C_{\kappa}$ is the angular convergence power spectrum adjusted for the lensing efficiency based on $z_l$ and $z_s$, 
\begin{equation}
\label{eq:convergen_mag}
 C_{\kappa}(\ell; z_l, z_s) = \int {\rm d} \chi\frac{W(\chi, \chi_l) W(\chi, \chi_s)}{\chi^2} P_{\rm mm}^{\rm NL}\left(\frac{\ell + 1/2}{\chi};z_l\right),
\end{equation}
where $W(\chi, \chi(z))$ is the lensing efficiency, defined by the moving distances $\chi$ and $\chi(z)$,
\begin{equation}
\label{eq:lensing_efficiency}
W(\chi, \chi(z)) = \frac{3}{2} \Omega_m H_0^2 (1+z) \frac{\chi (\chi(z) - \chi)}{\chi(z)}.
\end{equation}

\subsection{\label{sec:model:systematics} Source Systematics Error}

In this section, we will describe our modeling of the systematic errors in this work, including the redshift distribution uncertainties (Section~\ref{sec:model:nz_bias}) and multiplicative bias (Section~\ref{sec:model:multiplicative_bias}).

Note that intrinsic alignments (IA) of source galaxy shapes can in principle contaminate galaxy-galaxy lensing measurements via correlations between the lens positions and source galaxy intrinsic shapes \cite{Hirata2004}. However, previous studies (e.g., \cite{Mandelbaum2005, hirata2007, singh2015}) have found that IA contamination is subdominant for galaxy-galaxy lensing at the redshifts and scales considered here. We defer the impact of IA on cosmology inference based on tomographic galaxy-galaxy lensing to the $3\times 2$pt study.  Furthermore, we do not expect severe IA contamination to our GGL data vector, because our lens-source pair weights $w_{\rm ls}$ (Eq.~\ref{eq:sigma_inv_ls}) effectively downweight the lens-source pair that are physically associated. 

We also do not include Baryonic physics in this $2\times2$pt analysis following previous analysis \cite{Sugiyama2022}. In Section~\ref{sec:validation:validation}, we validate that the baryonic effect is neglible to the $2\times2$pt data vector. 

\subsubsection{\label{sec:model:nz_bias} Source Redshift Distribution Uncertainty}

As shown in the upper panel of Fig.~\ref{fig:redshift_distro}, our default redshift distribution $n_i(z)$ for the $i$-th tomographic bin is constructed by combining the photometric redshift probability distributions with clustering-based redshift calibration, following \cite{Rau2022}.

Following previous DES and HSC analyses \cite{Miyatake2022, Miyatake2023, Sugiyama2023, Faga2025, Zhang2023_nz}, the redshift distribution $n_i(z)$ for the $i$-th tomographic bin is modeled as a shifted version of the fiducial distribution, parameterized by a per-bin shift $\Delta z_i$: 
\begin{equation} 
n_i(z) \rightarrow n_i(z + \Delta z_i). 
\end{equation} 
The parameters $\Delta z_i$ allow for uncertainties in the mean redshift of each bin. For $\Delta z_1$ and $\Delta z_2$, we impose Gaussian priors based on the calibrated uncertainties from the full redshift distribution analysis in \cite{Rau2022}. However, in \cite{Li2023} and \cite{Dalal2023}, it was found that the $\Lambda$CDM model preferred values of $\Delta z_3$ and $\Delta z_4$ that were inconsistent with their photometric redshift calibrations. As a result, the informative priors on $\Delta z_3$ and $\Delta z_4$ were replaced with uniform priors over the range $[-1, 1]$ in those analyses.
In this work, we follow the same approach and adopt uninformative flat priors on $\Delta z_3$ and $\Delta z_4$. We can then test the ability to constrain these parameters using the tomographic galaxy-galaxy lensing measurements.

The correction to $\Delta \Sigma(R_p)$ due to shifts in the source redshift distribution is implemented by re-computing the lensing efficiency $\langle \Sigma_{\rm crit}^{-1} \rangle$ and the corresponding lensing weight $w_{\rm ls} = w_l w_s \langle \Sigma_{\rm crit}^{-1} \rangle^2_{\rm ls}$. We define a correction factor as 
\begin{equation}
\label{eq:redshift_correction}
f_{\Delta \Sigma}^{qi}(\Delta z_{i}) = \frac{\sum_{\rm l\in q, s\in i} w_{\rm l} w_{\rm s}  \langle \Sigma_{\rm crit}^{-1}\rangle^{\rm est}_{\rm ls} (\Delta z_i)}{\sum_{\rm l\in q, s\in i}  w_{\rm l} w_{\rm s}  \langle \Sigma_{\rm crit}^{-1}\rangle^{\rm fid}_{\rm ls}}.
\end{equation}
where the lens and source weight $w_l$ and $w_s$ are explained in Section~\ref{sec:measurement:ggl}. 
The corrected galaxy-galaxy lensing data vector for lens redshift $z_l$ and the $i$-th source bin is given by 
\begin{equation} 
\label{eq:corrected_dsigma} 
\Delta \Sigma^{qi}_{\rm corr}(R_p, \Delta z_i) = f_{\Delta \Sigma}^{qi}(\Delta z_i) \Delta \Sigma^{qi}(R_p). 
\end{equation} 
Here, $\Delta \Sigma(R_p | z_l)$ includes contributions from the galaxy-galaxy lensing signal $\Delta \Sigma_{gG}$, the magnification bias term $\Delta \Sigma_{\rm mag}$, and the 1-halo correction term $\Delta \Sigma_{\rm PM}$. 

We note that shifting the source redshift can theoretically change the boost factors that go into the GGL measurement, we leave  the modeling of redshift dependent boost factors to future studies.

\subsubsection{\label{sec:model:multiplicative_bias} Multiplicative bias}

The multiplicative bias of the HSC Y3 shear catalog is calibrated to the 1\% level using image simulations \cite{Li2022}. To account for residual uncertainty in this calibration, we include a 1\% multiplicative uncertainty in the galaxy-galaxy lensing model for each source bin, parameterized by a nuisance parameter $\Delta m_i$. The lensing signal is then modified as: 
\begin{equation} 
\label{eq:multiplicative_bias_corr} 
\Delta \Sigma^{qi} (R_p) \rightarrow (1 + \Delta m_i) \Delta \Sigma^{qi} (R_p). 
\end{equation}

\subsection{\label{sec:model:bayesian} Bayesian Inference}

In this section, we summarize the Bayesian inference framework used in this work for the cosmological analysis. The parameter space and the priors adopted in the analysis are described in Section~\ref{sec:model:params}, while the likelihood function and the construction of the data vector are presented in Section~\ref{sec:model:likelihood}.

\subsubsection{\label{sec:model:params} Parameter Space and Prior}

As described in Sections~\ref{sec:model:theory} and \ref{sec:model:systematics}, we construct a forward model to generate theoretical data vectors for $w_p$ and $\Delta \Sigma$ based on a set of cosmological and nuisance parameters. The cosmological parameter space consists of five flat $\Lambda$CDM parameters: $[\Omega_m, \Omega_c h^2, \Omega_b h^2, n_s, \log(10^{10}A_s)]$. Among these, $\Omega_m$ and $\log(10^{10}A_s)$ are the parameters to which weak lensing is most sensitive; therefore, we adopt uninformative (flat) priors for them. Following \cite{Planck2018Cosmology}, we adopt a flat prior on $\log(10^{10} A_s)$. We note that recent works \cite{Lemos2022} have investigated the impact of different prior choices and found that, while the differences are generally small for stage III weak lensing analyses, they may become more significant in future analyses.

The galaxy-galaxy lensing and clustering signals are not strongly sensitive to $\Omega_b$ and $n_s$, so we apply informative priors based on external measurements. Specifically, we adopt a prior on $\omega_b = \Omega_b h^2$ from Big Bang Nucleosynthesis (BBN) constraints \cite{Schoneberg2019}, and a prior on $n_s$ from Planck \cite{Planck2018Cosmology}. We leave the prior on $\omega_c = \Omega_c h^2$ uninformative to allow greater flexibility in the derived Hubble constant $h$.

For each lens tomographic bin, we include three nuisance parameters: a linear galaxy bias parameter $b_l(z_l)$, a point-mass parameter $\Delta \Sigma_{\rm PM}(z_l)$  to model the 1-halo correction, and a magnification bias parameter $\alpha_{\rm mag}(z_l)$ to describe the amplitude of the magnification contribution $\Delta \Sigma_{\rm mag}(z_l)$. Both the galaxy bias and point-mass parameters are assigned uninformative (flat) priors. This is justified because the combination of $w_p$ and $\Delta \Sigma$ is expected to break the degeneracy between galaxy bias and the amplitude of the matter power spectrum, and $\Delta \Sigma(R_p)$ in the range $2 < R_p < 4\ h^{-1}\mathrm{Mpc}$ provides sensitivity to the point-mass term $\Delta \Sigma_{\rm PM}(z_l)$.

The prior on the magnification bias parameter $\alpha_{\rm mag}(z_l)$ is Gaussian, with the mean estimated from the slope of the galaxy number counts at the luminosity threshold of each lens sample. We adopt a conservative standard deviation of 0.5 to account for systematic uncertainty in this estimate. In total, we include 9 nuisance parameters associated with the lens catalog.

For each source tomographic bin, we include two nuisance parameters: one for the multiplicative shear bias, $\Delta m_i$, and one for the shift in the source redshift distribution, $\Delta z_i$. The priors on the multiplicative bias parameters $\Delta m_i$ are Gaussian with zero mean and a standard deviation of $\sigma = 0.01$. The priors on $\Delta z_1$ and $\Delta z_2$ are also Gaussian, with zero mean and standard deviations estimated from the full redshift distribution uncertainty in \cite{Rau2022}. For $\Delta z_3$ and $\Delta z_4$, we adopt uninformative (flat) priors, allowing us to assess the constraining power on the source redshift distribution using tomographic galaxy-galaxy lensing over a wide range of scales.

In total, we include 8 nuisance parameters associated with the source tomographic bins. A summary of all model parameters and their priors in our fiducial analysis is provided in Table~\ref{tab:prior}, which includes 5 cosmological parameters and 17 nuisance parameters.

\begin{table}\centering
\begin{tabular}{ccc}
\hline
Parameter          &   & Prior          \\ \hline
$\log(A_s\times 10^{10})$   & \hspace{2cm}  & $U[1.0,5.0]$            \\
$\omega_b$          &    & $\mathcal{N}(0.02268,0.00038)$      \\
$n_s$               &   & $\mathcal{N}(0.9646,0.0126)$    \\
$\omega_c$          &     & $U[0.0998, 0.1398]$     \\
$\Omega_m$          &  & $U[0.0906, 0.5406]$    \\
$\tau$              &  & $0.0561$       \\
$\Omega_\nu$       &   & $0.06$         \\
$\Omega_k$              &   & $0$    \\
$w$                &   & $-1.0$    \\
$w_a$              &   & $0.0$    \\\hline
$b_1$              &   & $U[0.1,5.0]$ \\
$b_2$              &   & $U[0.1,5.0]$  \\
$b_3$              &   & $U[0.1,5.0]$  \\
$\Delta \Sigma_{\rm PM,1} (4 {\rm Mpc/h})$ &   & $U[0.0,10.0]$            \\
$\Delta \Sigma_{\rm PM,2} (4 {\rm Mpc/h})$ &  & $U[0.0,10.0]$              \\
$\Delta \Sigma_{\rm PM,3} (4 {\rm Mpc/h})$ &   & $U[0.0,10.0]$              \\
$\alpha_{\rm mag, 1}$    &    & $\mathcal{N}(2.258,0.5)$          \\
$\alpha_{\rm mag, 2}$    &    & $\mathcal{N}(3.563,0.5)$         \\
$\alpha_{\rm mag, 3}$    &    & $\mathcal{N}(3.729,0.5)$         \\\hline
$m_1$       &    & $\mathcal{N}(0.0,0.01)$  \\
$m_2$       &    & $\mathcal{N}(0.0,0.01)$ \\
$m_3$       &    & $\mathcal{N}(0.0,0.01)$ \\
$m_4$       &    & $\mathcal{N}(0.0,0.01)$  \\
$\Delta z_{1}$    &    & $\mathcal{N}(0,0.024)$    \\
$\Delta z_{2}$     &      & $\mathcal{N}(0,0.022)$      \\
$\Delta z_{3}$     &     & $U[-1,1]$     \\
$\Delta z_{4}$     &    & $U[-1,1]$   \\\hline
\end{tabular}
\caption{Summary of the model parameters and their priors used in the fiducial cosmological analysis. The top block lists the cosmological parameters, including five free flat-$\Lambda$CDM parameters and fixed values for $\tau$, $\Omega_\nu$, $\Omega_k$, $w$, and $w_a$. The middle block includes nuisance parameters associated with the lens galaxies: the linear galaxy biases ($b_1$, $b_2$, $b_3$), point-mass corrections for the 1-halo term at $R_0 = 4\ h^{-1}\mathrm{Mpc}$, and magnification bias parameters. The bottom block contains nuisance parameters for the source galaxies: multiplicative shear bias terms ($m_i$) and redshift shift parameters ($\Delta z_i$). Flat priors ($U[a,b]$) are uninformative, while Gaussian priors ($\mathcal{N}(\mu, \sigma)$) incorporate external constraints.
}
\label{tab:prior}
\end{table}

\subsubsection{\label{sec:model:likelihood} Likelihood Function}

The cosmological constraints in this work are obtained through Bayesian inference, which samples the parameter space described in Section~\ref{sec:model:params}, computes a theoretical data vector $\mathbf{d}^{\rm theory}$ for each parameter set $p$, and compares it with the observed data vector $\mathbf{d}^{\rm obs}$. The combined data vector $\mathbf{d}$ includes contributions from both galaxy clustering ($w_p$) and galaxy-galaxy lensing ($\Delta \Sigma$), and is structured as follows: 
\begin{equation} 
\mathbf{d} = { w_{p,1}, w_{p,2}, w_{p,3}, \Delta \Sigma_{11}, \Delta \Sigma_{12}, \dots, \Delta \Sigma_{34} }. 
\end{equation} 
There are 3 tomographic bin pairs for the clustering signal $w_{p,l}$ and 10 tomographic bin pairs for the galaxy-galaxy lensing signal. These 10 pairs arise from 3 lens bins and 4 source bins, excluding CMASS1–HSC1 and CMASS2–HSC1 due to significant redshift overlap.

For each $w_p$ bin pair, the data vector contains 14 data points logarithmically spaced over the range $[8, 80]\ h^{-1}\mathrm{Mpc}$. The lower scale cut at $8\ h^{-1}\mathrm{Mpc}$ is chosen to exclude scales dominated by nonlinear clustering effects \cite{Sugiyama2023}, while the upper scale cut avoids scales influenced by Baryon Acoustic Oscillations (BAO). In total, the clustering data vector includes 42 data points.

For each bin pair of $\Delta \Sigma$, we include 14 data points spaced logarithmically within the range $[2, 70]\ h^{-1}\mathrm{Mpc}$ in the analysis. Thanks to the point-mass correction term, we are able to model the small-scale behavior of $\Delta \Sigma$ down to $2\ h^{-1}\mathrm{Mpc}$ using the minimal bias model, as described in Section~\ref{sec:model:theory_ggl}.

The minimum scale cut is determined based on mock validation analyses that included scenarios with off-centering of lens galaxies, which is further discussed at Section~\ref{sec:res:offcentering}. We find that lowering the minimum scale from $2\ h^{-1}\mathrm{Mpc}$ to $1.5\ h^{-1}\mathrm{Mpc}$ introduces a significant bias in $S_8$, and thus we conservatively set the lower limit at $2\ h^{-1}\mathrm{Mpc}$. 

The maximum scale cut for $\Delta \Sigma(R_p)$ is determined from the cross term of the shear in the null test described in Section~\ref{sec:measurement:null}. Specifically, we exclude the largest-scale data point for each lens–source pair, as its inclusion leads to a significant decrease in the $p$-value of the cross term null test.
There are 140 data points used for the galaxy-galaxy lensing signal.

The total data vector, consisting of 182 elements, is measured on three blinded catalogs as well as on the 1404 mock catalogs described in Section~\ref{sec:data:mock}, which are used to estimate the covariance matrix $\mathbf{C}$. The log-likelihood function for a given parameter set $p$ is defined as 
\begin{equation} 
\label{eq:log_likelihood} 
\log \mathcal{L}(\mathbf{d} | p) = -\frac{1}{2} [\mathbf{d}^{\rm obs} - \mathbf{d}^{\rm theory}(p)]^{T} \mathbf{C}^{-1} [\mathbf{d}^{\rm obs} - \mathbf{d}^{\rm theory}(p)]. 
\end{equation} 
Here, we assume a multivariate Gaussian distribution for the uncertainties in the data vector, and $\mathbf{C}^{-1}$ denotes the inverse of the covariance matrix, corrected by the Hartlap factor \cite{Hartlap2007} when computing the likelihood.

To properly account for the uncertainties introduced by magnification bias, the theoretical magnification signal factor is multiplied with the mock measurements prior to computing the covariance matrix, since the magnification effect is not simulated in the mocks. This ensures that the resulting covariance accurately reflects the total uncertainty present in the observed data vector.

We use the posterior distribution over the parameter space to derive cosmological constraints from the SDSS and HSC Y3 datasets. The log-posterior is proportional to the sum of the log-prior and log-likelihood: 
\begin{equation} 
\label{eq:log_posterior} 
\log \mathrm{Post}(p | \mathbf{d}) \propto \log \mathrm{Prior}(p) + \log \mathcal{L}(\mathbf{d} | p). 
\end{equation} 
The priors adopted in this work are described in Section~\ref{sec:model:params}.
We sample the posterior distribution using a Monte Carlo method suitable for high-dimensional parameter spaces. Specifically, we employ the nested sampling algorithm \textsc{MultiNest}, as implemented in \textsc{CosmoSIS} \cite{Zuntz2015}. For the fiducial analysis and internal consistency tests, we set the \texttt{nlive} parameter to 500, the \texttt{efficiency} parameter to 0.3, and the \texttt{tolerance} parameter to 0.3.

Following previous HSC analysis \cite{Li2023,Dalal2023}, we report the summary statistics of the fiducial Bayesian inference in the following format: 
\begin{equation} 
\mathrm{mode}^{+34\%, \mathrm{upper}}_{-34\%, \mathrm{lower}} ({\rm MAP, mean}), 
\end{equation} 
where the mode is defined as the maximum of the one-dimensional marginal posterior distribution, estimated by interpolating the one-dimensional histogram. The $1\sigma$ uncertainty on each parameter is given by the 16th and 84th percentiles of the marginal posterior distribution.
We do not perform a Maximum A Posteriori (MAP) search for every analysis due to computing cost, but we report the MAP value for the fiducial analysis. 

\section{\label{sec:validation:0} Validation and Internal Consistency Tests}

To ensure that the analysis yields unbiased and robust cosmological results, we perform both validation tests and internal consistency tests for the 2$\times$2pt analysis. In the validation tests, we run the fiducial analysis settings on a variety of synthetic data vectors that incorporate different sources of systematics, verifying that our results are not biased by unaccounted-for systematic effects. In the internal consistency tests, we analyze different configurations of the real data that should yield statistically consistent results, thereby validating the robustness of the analysis pipeline. Additionally, we split the dataset by different lens–source bin pairs and survey footprints to check for any significant redshift-dependent or footprint-dependent biases.

\subsection{\label{sec:validation:validation} Validation Tests}

To validate that our model yields unbiased cosmological results, we perform analyses on noiseless synthetic data vectors generated from simulations. These simulations populate SDSS galaxies within dark matter halos identified in $N$-body simulations, and were previously used to validate earlier HSC analyses \cite{Sugiyama2022, Miyatake2022b, Sugiyama2023, Miyatake2023}.

The validation suite includes synthetic data vectors constructed under a variety of scenarios: different methods for populating satellite galaxies within host halos, off-centering of central galaxies, inclusion of baryonic feedback, strong and extreme assembly bias effects, incompleteness of central galaxy identification, and galaxy assignment based on friend-of-friend halos.

We also test the impact of redshift calibration by running the fiducial analysis on theoretical data vectors with significant redshift bias. These tests demonstrate that our treatment of redshift priors is essential to avoid biased cosmological inferences.

The validation tests show that the tomographic 2$\times$2pt analysis with point-mass correction yields $S_8$ values that are within $0.5\sigma$ of the true value used to generate the $N$-body simulations in all cases, except for the scenario with extremely large assembly bias, where the inferred $S_8$ is biased low by $0.67\sigma$. This demonstrates that our modeling framework produces unbiased $S_8$ constraints under a wide range of systematic variations, except in cases where the assembly bias is unrealistically large.

We also observe that the model systematically produces $\Omega_m$ values that are higher than the true input value, with the bias typically exceeding the $0.5\sigma$ level. This trend is consistent with the findings of the consistency tests reported in \cite{Sugiyama2023}. As a result, we do not report $\Omega_m$ constraints as a key finding of this work.

The full list of validation tests that are conducted in this work are shown in Table~\ref{tab:validation}. We show the results of these validation tests in Section~\ref{sec:results:validation}.

\begin{table*}
\setlength{\tabcolsep}{20pt}
\begin{center}
\begin{tabular}{ll}
\hline\hline
Mock test name & Description\\
\hline
\multicolumn{2}{l}{\hspace{-1em}{\bf Fiducial mock analyses}}\\
Fiducial Mocks                     & Fiducial mock data vector measured on simulation\\
\hline
\multicolumn{2}{l}{\hspace{-1em}{\bf Effects of galaxy population and location}}\\
nonfidNsat, sat-DM, sat-subhalo
        & Populate the satellite galaxies in the halos with different models\\
off centering 1, 2, 3, 4   & Four mocks realizations of shifting the positions of the central galaxies from the halo centers \\ 
assembly bias              & Mocks with large assembly bias \\
baryon                     & Mocks with baryonic feedback effect\\
incompleteness             & Mocks with incomplete central galaxies \\
fof                        & Galaxies populated with friend-of-friend halos \\ 
\hline
\multicolumn{2}{l}{\hspace{-1em}{\bf Parameters for the point-mass correction model}}\\
$R_{p,\rm min} = 1.5 h^{-1}$Mpc,  $R_{\rm 0} = 3 h^{-1}$Mpc,                                         &  Aggressive lower limit on $R_{p,\rm min}$ and $R_0$\\
$R_{p,\rm min} = 3 h^{-1}$Mpc,  $R_{\rm 0} = 6 h^{-1}$Mpc,                                         &  Conservative lower limit on $R_{p,\rm min}$ and $R_0$\\ 
\hline
\multicolumn{2}{l}{\hspace{-1em}{\bf Photo-$z$ bias on $\Delta \Sigma$}}\\
$\Delta z_{3,4} = 0.0$, $\Delta z_{3,4} \sim \mathcal{N}(0,\sigma_{z,3,4})$       & No redshift bias in the mocks, set the prior on $\Delta z_{3,4}$to informative priors provided by \cite{Rau2022}\\
$\Delta z_{3,4} = 0.2$, $\Delta z_{3,4} \sim \mathcal{N}(0,\sigma_{z,3,4})$       & Use $\Delta z_3=\Delta z_4=-0.2$ to compute $\Delta \Sigma$, use informative prior centered on $0$\\
$\Delta z_{3,4} = 0.2$, $\Delta z_{3,4} \sim U[-1,1]$       & $\Delta z_3$ and $\Delta z_4$ are set to -0.2, use the uninformative prior $ \sim U[-1,1]$. \\

\hline\hline
\end{tabular}
\end{center}
\caption{\label{tab:validation} Summary of validation tests conducted in this work. The mock data vector column shows the name of the different data vectors used for the validation tests, followed by a brief description of their difference to the fiducial mocks. We refer to \cite{Miyatake2022} for a detailed description of the simulation that generated these mock datasets. 
}
\end{table*}

\subsection{\label{sec:validation:internal_consistency}Internal Consistency Tests}

\begin{table*}
\renewcommand{\arraystretch}{1.2}
\setlength{\tabcolsep}{15pt}
\begin{center}
\begin{ruledtabular}
\begin{tabular}{llc}
Internal consistency test name & Description & $\mathcal{D}(\bm{p})$, $\mathcal{D}(\bm{d})$\\
\hline
Fiducial Analysis & fiducial $2\times2$pt analysis from section~\ref{sec:model:bayesian}                                                         & 22, 182\\
\hline
2$\times$2pt, w/o LOWZ$  $                                 & without LOWZ sample for $\Delta \Sigma$ and $w_p$                                     & 19, 126\\
2$\times$2pt, w/o CMASS1$  $                               & without CMASS1 sample for $\Delta \Sigma$ and $w_p$                                       & 19, 112\\
2$\times$2pt, w/o CMASS2$  $                               & without CMASS2 sample for $\Delta \Sigma$ and $w_p$                                     & 19, 112\\
\hline
2$\times$2pt, w/o $z_1$                                   &  without $1$st source bin for $\Delta \Sigma$                                          & 20, 168\\
2$\times$2pt, w/o $z_2$                                 &  without $2$nd source bin for $\Delta \Sigma$                                          & 20, 140\\
2$\times$2pt, w/o $z_3$                                  & without $3$rd source bin for $\Delta \Sigma$                                          & 20, 140\\
2$\times$2pt, w/o $z_4$                                  &  without $4$th source bin for $\Delta \Sigma$                                          & 20, 140\\
\hline
no photo-$z$ uncertainty                                               &  fixing $\Delta z_{i} = 0$                                                            & 18, 182\\
no shear uncertainty  & 2$\times$2pt, fixing $\Delta m_i = 0$  & 18, 182\\
no magnification bias uncertainty${}$                             &  fixing $\alpha_{\rm mag}=\mu$                                                            & 19, 182\\
\hline
XMM      ($33.17~{\rm deg}^2$)$  $                           &   using only XMM field for $\Delta \Sigma$                                    & 22, 182\\
GAMA15H  ($40.87~{\rm deg}^2$)$  $                           &   using only GAMA15H field for $\Delta \Sigma$                                 & 22, 182\\
HECTOMAP ($43.09~{\rm deg}^2$)$  $                           &  using only HECTOMAP field for $\Delta \Sigma$                                & 22, 182\\
GAMA09H  ($78.85~{\rm deg}^2$)$  $                           &   using only GAMA09H field for $\Delta \Sigma$                                & 22, 182\\
VVDS     ($96.18~{\rm deg}^2$)$  $                           &   using only VVDS field for $\Delta \Sigma$                                    & 22, 182\\
WIDE12H  ($121.32~{\rm deg}^2$)$  $                          &   using only WIDE12H field for $\Delta \Sigma$                               & 22, 182\\
\hline
DNNz  &  using DNNz for lensing measurement, and DNNz stacked $n(z)$                                            & 22, 182\\
Mizuki &using Mizuki for lensing measurement, and Mizuki stacked $n(z)$          & 22, 182\\
DEmPz                                         & 
using DEmPz for lensing measurement, and DEmPz stacked $n(z)$   & 22, 182
\end{tabular}
\end{ruledtabular}
\end{center}
\caption{\label{tab:internal_consistency} Summary of internal consistency tests ran in this work. Besides the fiducial analysis ran on the fiducial data vector described in Section~\ref{sec:model:likelihood} on parameter space described in Section~\ref{sec:model:params}, we split the datasets, apply different prior on the nuisance parameters, and use different photo-$z$ methods to quantify the robustness of the analysis. The last column displays the dimensionality of the parameter space and the data vector space. }
\renewcommand{\arraystretch}{1}
\end{table*}

In addition to our fiducial analysis, we perform a series of internal consistency tests to ensure the robustness of our results. These tests verify that there are no significant redshift-dependent or footprint-dependent trends in the inferred cosmological parameters. We also test the impact of different photometric redshift methods used to compute the optimal lens–source weighting $w_{\rm ls}$ and the source redshift distributions. We find that variations in photo-$z$ methodology do not significantly affect the cosmological results.

Table~\ref{tab:internal_consistency} summarizes all internal consistency tests performed to validate our results. The fiducial analysis uses the full range of $w_p$ measurements across three tomographic lens bin pairs and $\Delta \Sigma$ measurements across ten tomographic lens–source bin pairs, applying the fiducial scale cuts. 

The fiducial analysis incorporates photo-$z$ uncertainties, shear multiplicative bias, and magnification bias uncertainties. It uses the DNNz photometric redshift method for computing galaxy-galaxy lensing, and models the redshift distribution using the DNNz-clustering-calibrated $n(z)$.

In addition to the fiducial analysis, we perform a series of tests by splitting the dataset. First, we run analyses with data vectors that exclude one lens bin at a time. In each case, the clustering signal $w_p$ and all $\Delta \Sigma$ bin pairs associated with the excluded lens bin are removed from the data vector. Correspondingly, the parameter space is reduced by three because of the removal of the associated galaxy bias, magnification bias uncertainty, and point-mass correction parameter.

We also conduct analyses excluding one source bin at a time. In these cases, all $\Delta \Sigma$ bin pairs involving the excluded source bin are removed from the data vector, and two nuisance parameters are dropped from the model: the residual multiplicative bias and the source redshift shift parameter associated with that source bin.

We also perform a set of analyses in which we use only one of the six HSC fields at a time for the source sample in the galaxy-galaxy lensing measurement. For each field-specific analysis, we recompute the covariance matrix of the galaxy-galaxy lensing data vector using the corresponding HSC field from the mock catalogs, ensuring that the statistical uncertainties are properly estimated for the reduced footprint.

We also perform analyses in which we fix each class of systematic uncertainty that is marginalized over in the fiducial analysis. Specifically, the “no photo-$z$ uncertainty” and “no shear uncertainty” runs reduce the dimensionality of the parameter space by four, while the “no magnification bias uncertainty” run reduces it by three. It is important to note that, due to the empirical evidence for a non-negligible photo-$z$ bias in the third and fourth source bins, we expect the results of the “no photo-$z$ uncertainty” analysis to be inconsistent with those from the fiducial analysis.

For robustness, we also examine the impact of the choice of photometric redshift method on our results. Specifically, we repeat the analysis using DNNz, Mizuki, and DEmPz for both the galaxy-galaxy lensing measurement and the corresponding stacked source redshift distributions $n(z)$ used in the modeling. Note that the covariance matrix is always computed with DNNz to reduce the computational cost. 

In Section~\ref{sec:results:internal_consistency}, we present the outcomes of the internal consistency tests. For each test, we compare the mode of the $S_8$ posterior with that from the fiducial analysis. 
% We consider the results consistent if the shift in $S_8$ is within $0.5\sigma$, where $\sigma$ is the posterior uncertainty on $S_8$. 
For the single-field runs, we use the larger of the $\sigma$ values (between the single-field and fiducial analyses) when assessing the significance of any shift, to provide a conservative estimate. We exclude the “no photo-$z$ bias” test from this consistency assessment, since that analysis setup is intentionally unrealistic.

\section{\label{sec:results:0} Cosmological Constraints}

In this section, we first present the cosmological constraints obtained from our fiducial analysis, in Section~\ref{sec:results:fiducial}.
Next, we discuss the outcomes of the internal consistency tests, which are used to assess the robustness of our methodology, in Section~\ref{sec:results:internal_consistency}.
Last but not least, the results of the validation tests using synthetic data vectors are provided in Section~\ref{sec:results:validation}.

\subsection{\label{sec:results:fiducial} Fiducial Constraints}

\begin{figure}
\includegraphics[width=0.95\columnwidth]{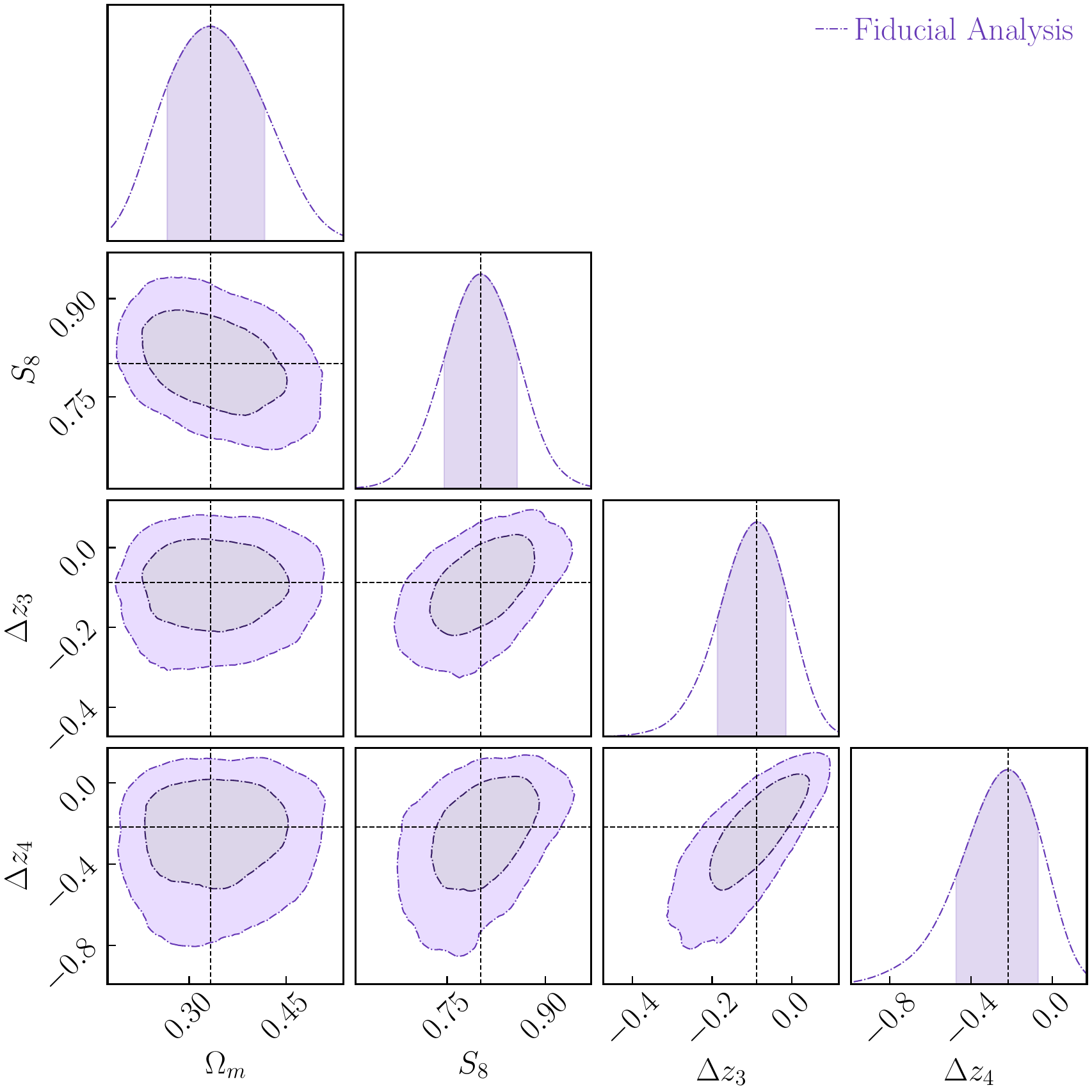}
\caption{\label{fig:fiducial_analysis} Posterior constraints from the fiducial $2\times2$pt analysis. The 68\% (1$\sigma$) and 95\% (2$\sigma$) credible contours are shown for the joint posterior distributions of $\Omega_m$, $S_8$, $\Delta z_3$, and $\Delta z_4$. The diagonal panels show the marginalized one-dimensional posterior distributions. The vertical dotted lines indicate the mode of each marginalized posterior. %\rachel{Marginalized over all other parameters?  And are the regions $1\sigma$ and $2\sigma$?} \tianqing{added some text} 
}
\end{figure}

 % After validating our pipeline using mock data vectors and performing blinded analyses—including both the fiducial and internal consistency tests across all three blinded catalogs—we confirm that the model proposed in this work can recover unbiased estimates of the $S_8$ parameter, while remaining robust against potential biases in the redshift distribution parameters.
Following the completion of all validation and consistency checks, we unblinded the results and identified the first catalog as the true dataset. In this section, we present the results of the fiducial analysis.

We show the 68\% (1$\sigma$) and 95\% (2$\sigma$) credible contours for the parameters $(\Omega_m, S_8, \Delta z_3, \Delta z_4)$ in Fig.~\ref{fig:fiducial_analysis}.
The reported values are presented in the format:
\begin{equation}
\nonumber \mathrm{mode}^{+34\%}_{-34\%} (\mathrm{MAP}, \mathrm{mean}),
\end{equation}
where the mode is the peak of the 1D marginalized posterior, MAP is the maximum {\em a posteriori} estimate, and the mean is the posterior mean.

Here are the summary statistics for primary parameter of interest:
\begin{align}
\mathbf{S_8} & \mathbf{ =0.804^{+0.051}_{-0.051} (0.797, 0.801)}\\
\mathbf{\Delta z_3} &\mathbf{= -0.079^{+0.074}_{-0.084} (-0.117, -0.101)}\\
\mathbf{\Delta z_4} &\mathbf{= -0.203^{+0.167}_{-0.206}(-0.263,-0.271)}.
\end{align}
The $S_8$ constraint from our fiducial analysis is statistically consistent with previous 2$\times$2pt results from \cite{Sugiyama2022, Miyatake2023}. Because the posterior distribution of $S_8$ is nearly symmetric, the mode, MAP, and mean are in close agreement, and the upper and lower 34\% credible intervals are nearly identical. The full set of posterior distributions from the fiducial analysis is shown in Appendix~\ref{ap:full_corner_plot}.

Our fiducial analysis demonstrates that the tomographic 2$\times$2pt framework can leverage the implicit shear ratio information in the galaxy-galaxy lensing data vector to calibrate the redshift distribution parameters $\Delta z_3$ and $\Delta z_4$. The central values obtained for these parameters are consistent with those derived from the cosmic shear analyses in \cite{Li2023, Dalal2023}, although the signal-to-noise ratios inferred from the errorbars in this work are lower than those achieved in the cosmic shear measurements.
We also observe that the marginalized posterior distributions of $\Delta z_3$ and $\Delta z_4$ are skewed toward zero, with the means lying further from zero than the modes. A similar trend was found in the validation tests using mock data vectors. From those tests, we find that the mode provides a more accurate point estimate of redshift bias than the mean, particularly when the posterior distribution is asymmetric.

Constraints on $\Omega_m$ and $\sigma_8$ are obtained but are subject to larger degeneracies and systematics in the 2×2pt analysis, and should be interpreted with caution
\begin{align}
\Omega_m &= 0.332^{+0.074}_{-0.069} (0.332, 0.341)\\
\sigma_8 &= 0.723^{+0.120}_{-0.095} (0.757, 0.765).
\end{align}
These constraints are also consistent with previous HSC findings. 

\begin{figure}
\includegraphics[width=0.95\columnwidth]{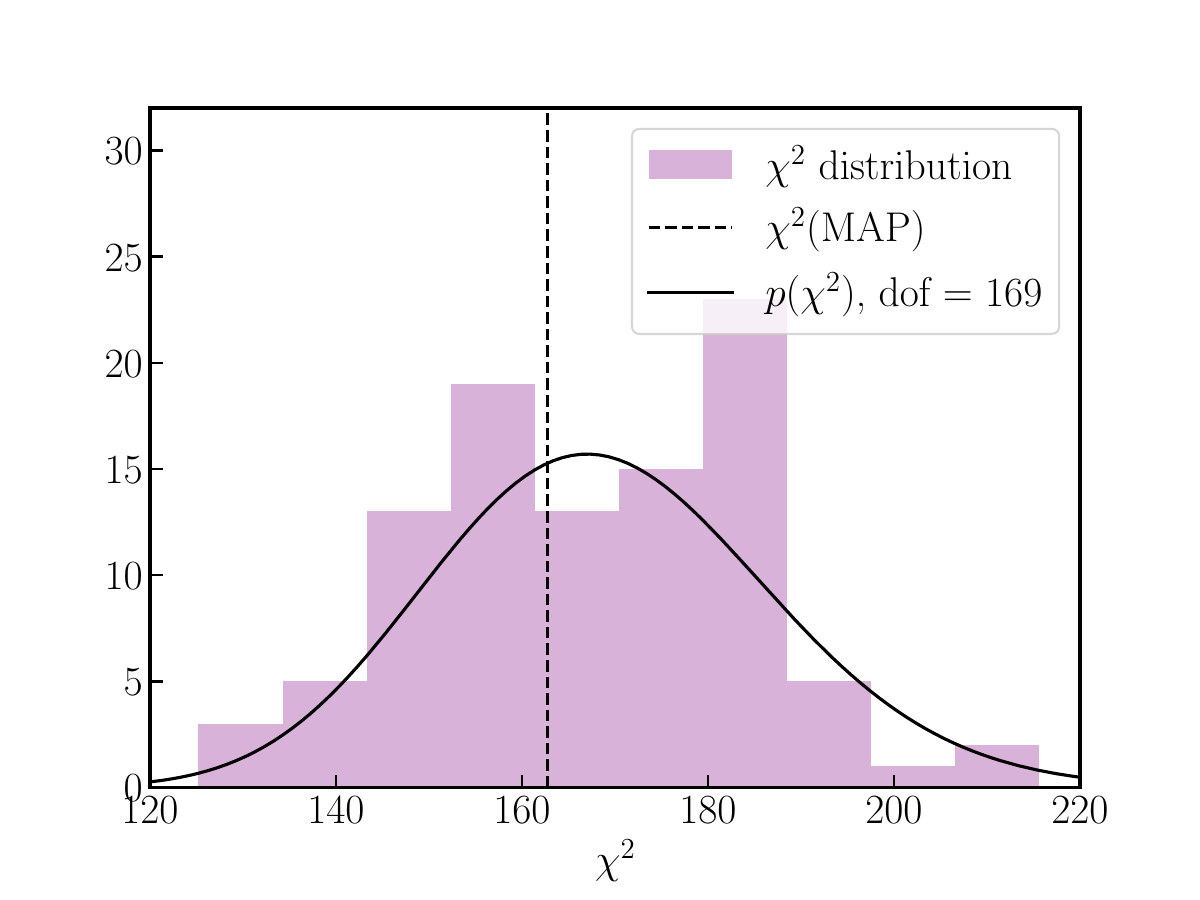}
\caption{\label{fig:chi2_dist} The purple distribution shows the MAP value of 100 noisy mock catalogs generated by our covariance matrix. The $\chi^2$ value of the MAP parameter of our fiducial run is shown in the black dotted line. With $\chi^2 ({\rm MAP}) = 162.7$, the p-value for the MAP parameter is $0.55$. }
\end{figure}

%need to describe the chi2 process

Fig.~\ref{fig:chi2_dist} shows the goodness-of-fit test for our fiducial analysis. To compute the effective $\chi^2$ distribution, we generate 100 noisy mock data vectors using the estimated covariance matrix $\tilde{\mathbf{C}}$, adjusted by the Hartlap factor \cite{Hartlap2007} to account for finite sample effects. For each mock realization, we run the MAP finder \texttt{Maxlike}, implemented in \texttt{CosmoSIS}, using the Nelder–Mead algorithm to identify the local parameter set that maximizes the posterior.
Because the MAP finder is sensitive to the choice of initial conditions, we repeat the optimization 10 times per mock realization with different starting points, and apply the same procedure to the real data vector. For our real dataset, the minimum $\chi^2$ is 162.7. The best-fit number of degrees of freedom (dof) for the effective $\chi^2$ distribution is 169, yielding a $p$-value of $p = 0.55$ for the fiducial analysis.

\subsection{\label{sec:results:internal_consistency} Internal Consistency Tests}

\begin{figure*}
\includegraphics[width=2\columnwidth]{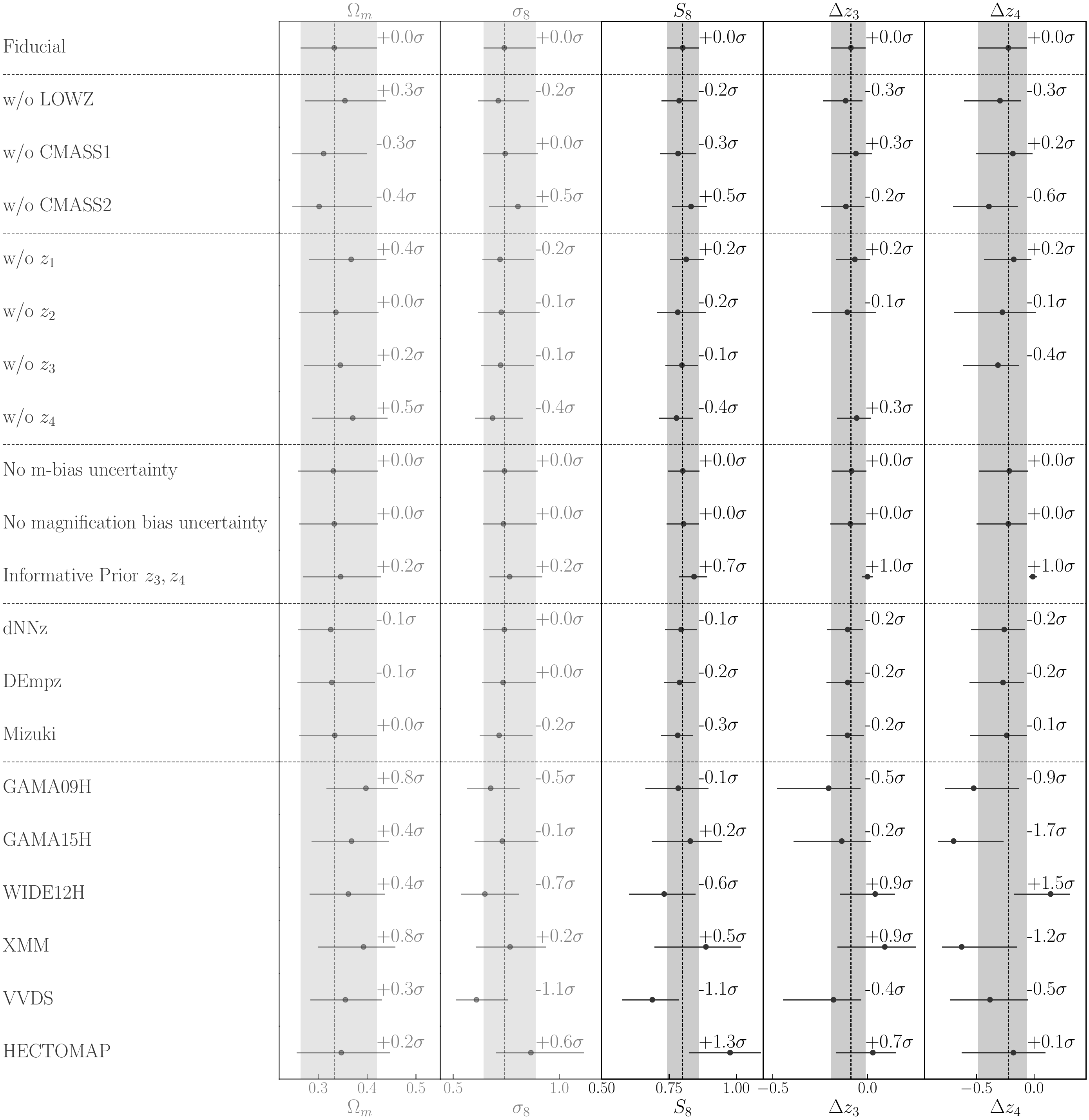}
\caption{\label{fig:internal_consistency} One-dimensional marginalized posterior constraints on five key parameters—-$\Omega_m$, $\sigma_8$, $S_8$, $\Delta z_3$, and $\Delta z_4$—-from the internal consistency tests described in Table~\ref{tab:internal_consistency}. Each point shows the posterior mode and its 68\% (1$\sigma$) credible interval. The vertical dotted lines indicate the modes from the fiducial analysis, and the shaded bands represent the 1$\sigma$ region of the fiducial constraints. The internal consistency tests use subsets or variants of the original data vector or deploy different priors on the nuisance parameters. No significant inconsistencies are found across these tests, indicating the robustness of the analysis. The higher priority parameters $S_8$, $\Delta z_3$, and $\Delta z_4$ are emphasized by darker shades. %\rachel{Refer to table III for more description of each scenario?  Say what types of data vectors were used for internal consistency tests?}\tianqing{Done.}
}
\end{figure*}

We perform extended internal consistency tests to evaluate the robustness of our analysis framework. These tests include running the analysis on subsets of the dataset and using alternative analysis settings, such as different redshift distributions or exclusion of specific lens/source bins. The internal consistency test configurations are summarized in Table~\ref{tab:internal_consistency}, and the resulting parameter constraints are presented in Fig.~\ref{fig:internal_consistency}.

Our first set of internal consistency tests removes one lens or source bin at a time to test for dependence on a particular tomographic bin. We do not observe a significant dependence of $S_8$ on the removal of any individual lens or source tomographic bin in our analysis. We note that the uncertainties on $\Delta z_3$ and $\Delta z_4$ increase the most when the second source bin is removed, indicating that self-calibration is anchored by the redshift distribution of the second source bin.

The second set of consistency tests involves assessing the sensitivity to the priors set on the nuisance parameters and different photo-$z$ algorithms. We fixed (removed the uncertainties on) the multiplicative shear bias and magnification bias parameters and found no significant change in the results. Most importantly, we performed an analysis using the priors on $\Delta z_3$ and $\Delta z_4$ from \cite{Rau2022}, i.e., $\Delta z_3 \sim \mathcal{N}(0, 0.031)$ and $\Delta z_4 \sim \mathcal{N}(0, 0.034)$. Similar to the findings in \cite{Li2023, Dalal2023, Sugiyama2023, Miyatake2023}, we obtain a higher $S_8$ value when using the informative $n(z)$ priors in the third and fourth bins. This further highlights the importance of adopting uninformative priors for the redshift distribution in our analysis.

We note that the parameter inferences from the first two sets of internal consistency tests are likely to be highly correlated with the fiducial analysis. [We refer readers to \cite{raveri2019} for detailed discussion] As a result, conclusions regarding whether observed parameter shifts are driven by statistical fluctuations cannot be made without quantifying the correlation between the inferences, ideally through dedicated simulations. Assuming two parameter estimates, $\boldsymbol{\theta}_1$ and $\boldsymbol{\theta}_2$, with identical uncertainties $\sigma$ and a correlation coefficient $\rho$, the standard deviation of $\boldsymbol{\theta}_1-\boldsymbol{\theta}_2$ under the assumption of Gaussian statistics is given by
\begin{equation}
\label{eq:correlation_metrics}
\sigma(\boldsymbol{\theta}_1 - \boldsymbol{\theta}_2) = \sigma \sqrt{2(1 - \rho)}.
\end{equation}
According to Eq.~\ref{eq:correlation_metrics}, a parameter shift of $0.5\sigma$ corresponds to a standard deviation between $\boldsymbol{\theta}_1$ and $\boldsymbol{\theta}_2$ with a correlation of $\rho = 0.875$, which is a high level of correlation. Therefore, we conclude that our requirement of agreement within $0.5\sigma$ is sufficiently stringent unless the analyses are extremely correlated.

Lastly, we conducted 2$\times$2pt analyses using only one HSC field at a time to examine whether our results exhibit significant field dependence. In these tests, both the galaxy-galaxy lensing data vector and its covariance are measured using data from a single field only. As a result, the constraining power is weaker compared to the fiducial analysis. We observe larger variance in $\sigma_8$, $S_8$, $\Delta z_3$, and $\Delta z_4$ across different fields than in $\Omega_m$, as these parameters are more sensitive to the amplitude of the lensing signal, which is weakly uncorrelated across fields.

Similar to the first two sets of tests, we expect the analyses based on different HSC fields to be correlated, primarily because they all share the full SDSS lens sample. However, we anticipate that the correlation in the $S_8$ constraints is low, as the $S_8$ measurement is primarily driven by the lensing data vector. This assumption is supported by the observation that the uncertainty on $S_8$ increases substantially when the lensing analysis is restricted to a single HSC field.

Overall, our internal consistency tests show no significant dependence of the cosmological or redshift distribution parameters on analysis choices. 
These assessments were performed on all three blinded catalogs prior to unblinding, ensuring that the final results are not influenced by analysis tuning or confirmation bias.

\subsection{Validation Tests}
\label{sec:results:validation}

\begin{figure*}
\includegraphics[width=2\columnwidth]{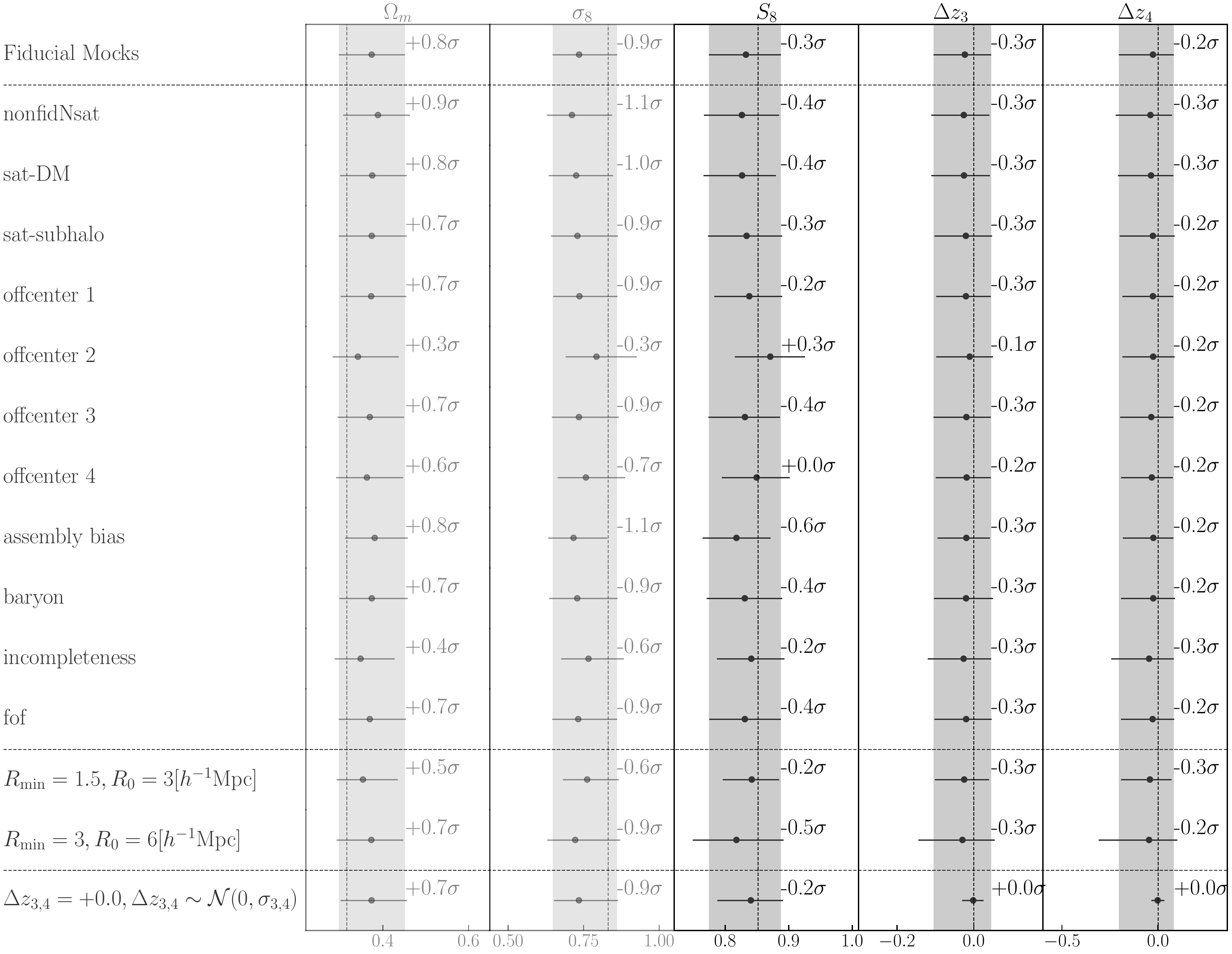}
\includegraphics[width=2\columnwidth]{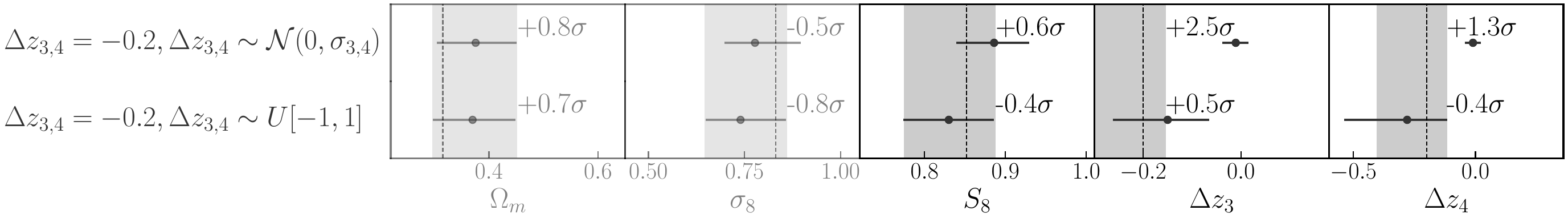}
\caption{\label{fig:validation} One-dimensional marginalized posterior constraints on five key parameters—$\Omega_m$, $\sigma_8$, $S_8$, $\Delta z_3$, and $\Delta z_4$—from the mock validation tests described in Table~\ref{tab:validation}. Each point shows the posterior mode with its 68\% (1$\sigma$) credible interval. The vertical dotted lines indicate the true input values from the simulations, and the shaded regions show the 1$\sigma$ constraints from the fiducial analysis on real data. Across all tests, we find no significant bias in the recovery of $S_8$, $\Delta z_3$, or $\Delta z_4$, confirming the robustness of the analysis framework against a variety of simulated systematics. The higher priority parameters $S_8$, $\Delta z_3$, and $\Delta z_4$ are emphasized by darker shades.  %\rachel{Refer to table II or a section for more description of each scenario?  Say what types of data vectors were used for mock validation tests?}\tianqing{Done}
}
\end{figure*}

In this section, we present the results of the validation tests, which are based on analyses conducted using mock lens and source catalogs generated from $N$-body simulations. The mock catalogs incorporate halo occupation distribution (HOD) modeling for populating lens galaxies and use ray-tracing techniques to compute lensing signals, as described in Section~\ref{sec:data:mock} and Section~\ref{sec:validation:validation}. The goal of these validation tests is to verify that our analysis pipeline yields unbiased constraints on cosmological and redshift distribution parameters, relative to the true input values used to generate the simulated datasets.

Fig.~\ref{fig:validation} shows the mode and 1$\sigma$ uncertainty of the one-dimensional marginalized posterior distributions for $\Omega_m$, $\sigma_8$, $S_8$, $\Delta z_3$, and $\Delta z_4$ across the validation tests listed in Table~\ref{tab:validation}. For the cosmological parameters, the dashed vertical lines indicate the true values used to generate the simulations. For $\Delta z_3$ and $\Delta z_4$, the dashed lines mark $\Delta z_{3,4} = 0.0$, which are the true values for all tests except the final two. The shaded bands represent the 1$\sigma$ uncertainty from the fiducial mock analysis. Comparing individual test cases to the fiducial result illustrates the impact of each systematic effect, while any discrepancy between the fiducial result and the true input parameters reflects a combination of model imperfections and projection effects \cite{Krause2010}.

\subsubsection{Lens galaxy population}

In the tests \texttt{nonfidNsat}, \texttt{sat-DM}, and \texttt{sat-subhalo}, we find no significant bias introduced by different methods of populating satellite galaxies within halos. Similarly, we observe no significant bias in the cosmological or redshift distribution parameters when the lens galaxy catalog differs from the fiducial mock in terms of assembly history, baryonic feedback, incompleteness, or galaxy-population methods. 

\subsubsection{Offcentering}
\label{sec:res:offcentering}

\begin{figure}
\includegraphics[width=1\columnwidth]{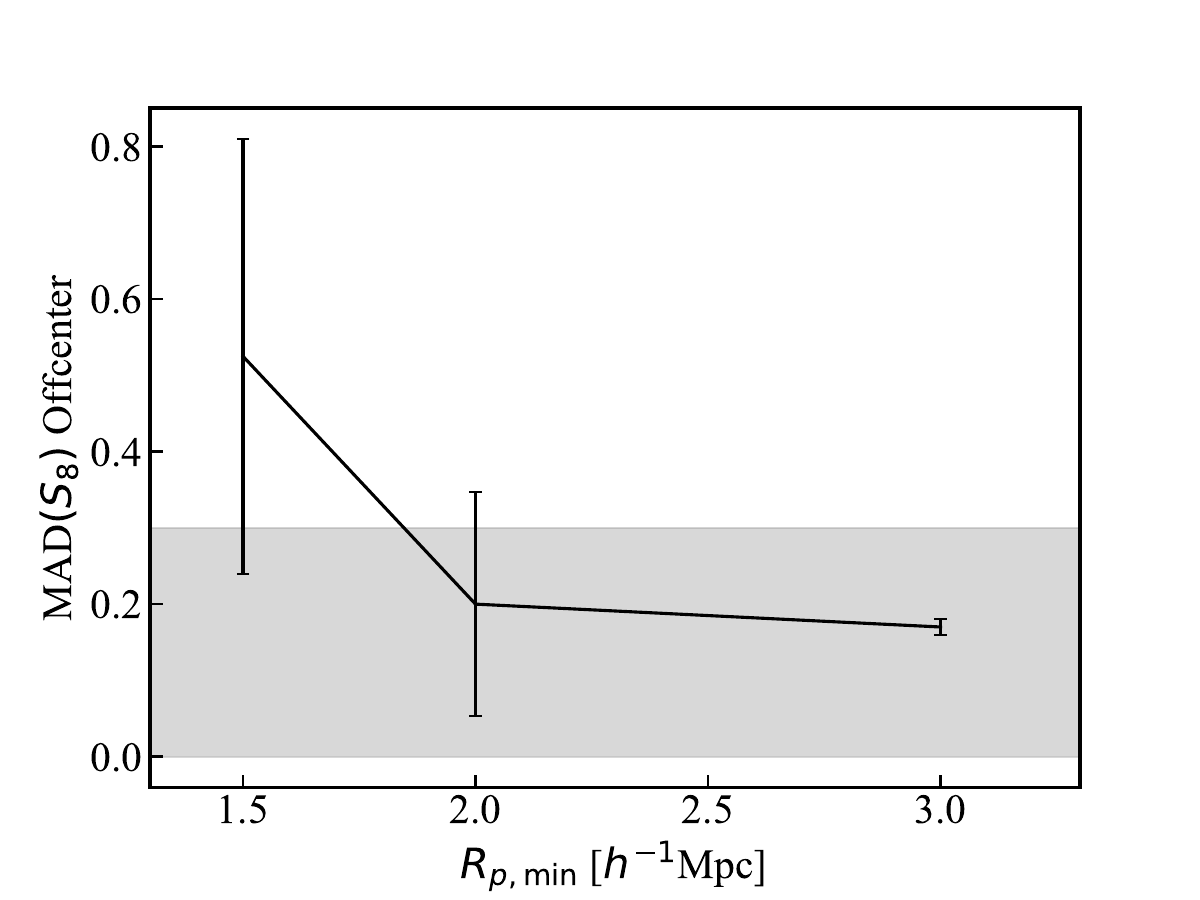}
\caption{\label{fig:rmin} The Median Absolute Deviation (MAD) of $S_8$ as a function of $R_{p,\rm min}$, estimated by 4 realizations of offcentering central galaxies. In each case, $R_0 = 2R_{p,\rm min}$ to ensure the same amount of data points used to estimate $\Upsilon(R_0)$. The optimal choice of $R_{p,\rm min}$ within the requirement in grey is $2 h^{-1}$Mpc, because it kept the $S_8$ bias under control while maintaining better constraining powers.  %\rachel{Elaborate on why?  Presumably both 2 and 3 are appropriate but theformer has better SNR?} \tianqing{2 over 3 is because of better SNR}
%\rachel{Need units in x axis. 
% Is 4 realizations enough to get believable errorbars?  It's weird how much they jump from 2 to 3.}\tianqing{The errorbar of the MAD is down to how sensitive the data vectors are to the offcentering. At 3 Mpc, I think the offcentering just barely makes a difference. I have added units. }
}
\end{figure}

For the four off-centering mock data vectors, we observe noticeable scatter in the inferred $S_8$ values. To quantify the sensitivity of $S_8$ to the choice of small-scale cuts, we run the fiducial and four off-centering mock data vectors using three different scale cut settings: (a) $R_{p,\rm min} = 1.5,h^{-1}$Mpc and $R_0 = 3,h^{-1}$Mpc; (b) $R_{p,\rm min} = 2,h^{-1}$Mpc and $R_0 = 4,h^{-1}$Mpc; and (c) $R_{p,\rm min} = 3,h^{-1}$Mpc and $R_0 = 6,h^{-1}$Mpc. Fig.~\ref{fig:rmin} shows the Median Absolute Deviation (MAD) of the $S_8$ results across these settings.

We find that the level of $S_8$ scatter depends on $R_{p,\rm min}$, the minimum projected separation used in the analysis. For our fiducial choice, setting (b), with $R_{p,\rm min} = 2,h^{-1}$Mpc, the MAD of $S_8$ is $0.2\sigma$, which is within acceptable limits. However, for setting (a), with $R_{p,\rm min} = 1.5,h^{-1}$Mpc, the MAD increases to $0.53\sigma$, indicating increased sensitivity to off-centering systematics.

Based on this, we adopt setting (b) —$R_{p,\rm min} = 2,h^{-1}$Mpc and $R_0 = 4,h^{-1}$Mpc — as our fiducial configuration, providing a favorable balance between signal-to-noise and robustness to small-scale modeling uncertainties.

\subsubsection{Redshift Distribution Parameters}
\label{sec:res:nz}

The last three validation tests demonstrate that our analysis can effectively self-calibrate redshift distribution biases in the source catalog. In the first test, with $\Delta z_{3,4} = 0.0$ and priors $\Delta z_{3,4} \sim \mathcal{N}(0, \sigma_{z,3,4})$, no redshift bias is introduced in the mock data vector, and the priors are set to the informative Gaussian distributions from \cite{Rau2022}. Compared to our fiducial analysis (which uses uninformative priors), the use of informative priors reduces the uncertainty on $S_8$ by only about 10%.

In the second test, with $\Delta z_{3,4} = -0.2$ and the same informative priors $\Delta z_{3,4} \sim \mathcal{N}(0, \sigma_{z,3,4})$, a significant redshift bias is introduced into the third and fourth HSC source bins. Since the informative priors remain centered at zero, this results in a $\sim1.2\sigma$ bias in the inferred $S_8$ compared to the fiducial analysis. This outcome is expected and highlights the motivation for avoiding informative priors on $\Delta z_{3,4}$ in our cosmological analysis.

In the final validation test, with $\Delta z_{3,4} = -0.2$ and priors $\Delta z_{3,4} \sim U[-1,1]$, we retain the source redshift bias of $-0.2$ but adopt uninformative priors instead of informative ones. The uninformative priors allow the posterior distributions of $\Delta z_{3,4}$ to re-center around the true values, resulting in an unbiased estimate of $S_8$ relative to the fiducial mock analysis. This test confirms that uninformative priors are essential for obtaining unbiased cosmological constraints when the source galaxy catalog may contain redshift distribution biases.

\section{\label{sec:conclusion:0} Summary and Discussion}

In this work, we performed a tomographic 2$\times$2pt analysis using galaxy clustering and galaxy-galaxy lensing measurements, with SDSS DR11 galaxies as lenses and HSC Y3 galaxies as sources. We adopted a point-mass correction model to extend the usable small-scale limit of the radial binning down to $2\,h^{-1}\mathrm{Mpc}$. To ensure that our measurements are not significantly affected by systematic errors, we conducted a series of validation tests, including mock analyses based on $N$-body simulations with HOD-populated lens galaxies. These tests also guided our choice of the small-scale cut to balance signal-to-noise and robustness to modeling uncertainties.

Our flat $\Lambda$CDM analysis yielded a constraint on the cosmological parameter $S_8 = 0.804^{+0.051}_{-0.051}$, consistent with previous HSC Y3 3$\times$2pt and cosmic shear studies. Importantly, it provides an independent validation of the redshift distribution parameters, with $\Delta z_3 = -0.079^{+0.074}_{-0.084}$ and $\Delta z_4 = -0.203^{+0.167}_{-0.206}$, in agreement with the results from the HSC Y3 cosmic shear analysis. We also computed the effective $\chi^2$ distribution for our analysis and found that the best-fitting model provides a good fit to the data vector, with a $p$-value of 0.54.

Our results demonstrate the self-calibration capability of galaxy-galaxy lensing in constraining the mean redshifts of tomographic source bins, particularly for the third and fourth bins, where previous HSC analyses adopted uninformative priors due to inconsistencies between $\Lambda$CDM predictions and photo-$z$ plus clustering redshift inferences. We anticipate that combining the 2$\times$2pt data vector with the cosmic shear measurements—using a consistent source bin definition—will allow the redshift calibration power of galaxy-galaxy lensing to directly benefit the cosmic shear constraints. This joint approach is expected to yield tighter cosmological constraints in a full 3$\times$2pt analysis. We refer the reader to \cite{3x2_paper} for the companion 3$\times$2pt analysis building on this work.

To ensure the robustness of our galaxy-galaxy lensing measurements, we conducted extensive null tests. These included examining the cross-component of the lensing signal, $\Delta \Sigma_{\times}$, which was found to be consistent with zero after subtracting the signal measured around random points, up to $70\,h^{-1}$Mpc. We also measured the boost factor, $B(R_p)$, to assess contamination from physically associated galaxies between the source and lens bins, as well as the lensing signal around random points to identify residual systematics. We found boost factors that exceed one for a subset of the lens-source redshift bin pairs, and multiply the boost factor by the lensing measurement to correct for the effect. These tests confirm that systematic effects in our measurements are well controlled, with no significant systematics detected within the scale range used for cosmological inference ($2$–$70\,h^{-1}$Mpc).

Our internal consistency tests confirmed that varying the photometric redshift methodology (DNNz, Mizuki, and DEmPz), modifying the tomographic bin combinations, altering priors on nuisance parameters other than the redshift distribution parameters $\Delta z_i$, and using different HSC fields do not significantly impact the final cosmological results. To validate that our analysis pipeline yields unbiased cosmological constraints, we performed validation tests using mock catalogs generated from $N$-body simulations. These tests demonstrate that our modeling choices—including the point-mass correction scheme, assumptions about galaxy population within dark matter halos, and the choice of redshift priors—do not introduce significant biases in $S_8$, $\Delta z_3$, or $\Delta z_4$. All analyses were performed on three blinded catalogs, preventing the results from confirmation bias.

This work improves the overall constraining power of the 2$\times$2pt analysis in three key ways compared to the previous HSC Y3 3$\times$2pt analysis:
(1) We include more source galaxies, particularly enhancing the signal in the LOWZ lens bins. Specifically, the total signal-to-noise ratio increases by up to 25\% for the LOWZ lens sample.
(2) We incorporate small-scale information in galaxy-galaxy lensing by applying the point-mass correction model, allowing the use of scales down to $2\,h^{-1}$Mpc.
(3) The inclusion of a lower-redshift source bin provides a crucial anchor for self-calibrating the redshift distribution in higher-redshift bins. This anchoring effect is evident from the increased uncertainty in $\Delta z_3$ and $\Delta z_4$ when the second HSC bin is excluded in the internal consistency tests.

Despite these advances, several challenges remain. First, while the point-mass correction model effectively mitigates small-scale mis-modeling, it introduces one additional nuisance parameter per lens tomographic bin, which can dilute the overall constraining power. Another limiting factor is the growth in the length of the data vector, which reduces the Hartlap factor and inflates the effective covariance matrix. This issue becomes even more significant in 3$\times$2pt analyses, where the length of the data vector may approach a substantial fraction of the number of simulations used to estimate the covariance. To address this, data compression techniques such as MOPED \cite{heavens2000} can be explored.
Lastly, given the modest self-calibration power of galaxy-galaxy lensing alone, further spectroscopic calibration efforts will be essential for the success of future weak lensing analyses. In particular, leveraging DESI spectra to refine the high-redshift tail of the source redshift distribution will be critical.

This work lays the foundation for an HSC Y3 $3\times 2$pt analysis using tomographic source galaxies, which is presented in \cite{3x2_paper}. \cite{3x2_paper} will focus more on comparing the results of this methodology to other surveys, and the implications for cosmology, while this work has focused on understanding and controlling systematic uncertainties in the method. Furthermore, The methodology developed in this work can be applied to upcoming $2\times2$pt analyses based on the combination of spectroscopic surveys like Dark Energy Spectrographic Instruments and weak lensing surveys such as the Rubin Observatory's LSST and Roman.

\begin{acknowledgments}

TZ and RM are supported by Schmidt Sciences. TZ thanks SLAC National Accelerator Laboratory for providing hospitality and an excellent research environment during the course of this study. RM is supported in part by a grant from the Simons Foundation (Simons Investigator in Astrophysics, Award ID 620789). AJN s supported by JSPS Kakenhi Grant Numbers: JP22K21349, JP23H00108 and JP25H0155. HM is supported by JSPS Kakenhi Grant Numbers: JP22K21349, JP23H00108, and JP24KK0065. MS is supported by JSPS Kakenhi Grant Numbers: JP24H00215 and JP24H00221. TS is supported by JSPS Kakenhi Grant Number: 24K17067. KO is supported by JSPS KAKENHI Grant Number JP24H00215, JP25K17380, JP25H01513, and JP25H00662. TN is supported by JSPS KAKENHI Grant Numbers: JP20H05861, JP23K20844, JP22K03634, JP24H00215, and JP24H00221. 

Work at Argonne National Laboratory was supported by the U.S. Department of Energy, Office of High Energy Physics. Argonne, a U.S. Department of Energy Office of Science Laboratory, is operated by UChicago Argonne LLC under contract no. DE-AC02-06CH11357. MMR acknowledges the Laboratory Directed Research and Development (LDRD) funding from Argonne National Laboratory, provided by the Director, Office of Science, of the U.S. Department of Energy under Contract No. DE-AC02-06CH11357. Work at Argonne National Laboratory was also supported
under the U.S. Department of Energy contract DE-AC02-06CH11357.

This work was supported in part by World Premier International Research Center Initiative (WPI Initiative), MEXT, Japan, and JSPS KAKENHI Grant Number 24H00215.

The Hyper Suprime-Cam (HSC) collaboration includes the astronomical communities of Japan and Taiwan, and Princeton University. The HSC instrumentation and software were developed by the National Astronomical Observatory of Japan (NAOJ), the Kavli Institute for the Physics and Mathematics of the Universe (Kavli IPMU), the University of Tokyo, the High Energy Accelerator Research Organization (KEK), the Academia Sinica Institute for Astronomy and Astrophysics in Taiwan (ASIAA), and Princeton University. Funding was contributed by the FIRST program from the Japanese Cabinet Office, the Ministry of Education, Culture, Sports, Science and Technology (MEXT), the Japan Society for the Promotion of Science (JSPS), Japan Science and Technology Agency (JST), the Toray Science Foundation, NAOJ, Kavli IPMU, KEK, ASIAA, and Princeton University.

This paper makes use of software developed for Vera C. Rubin Observatory. We thank the Rubin Observatory for making their code available as free software at http://pipelines.lsst.io/.

This paper is based on data collected at the Subaru Telescope and retrieved from the HSC data archive system, which is operated by the Subaru Telescope and Astronomy Data Center (ADC) at NAOJ. Data analysis was in part carried out with the cooperation of Center for Computational Astrophysics (CfCA), NAOJ. We are honored and grateful for the opportunity of observing the Universe from Maunakea, which has the cultural, historical and natural significance in Hawaii.

The Pan-STARRS1 Surveys (PS1) and the PS1 public science archive have been made possible through contributions by the Institute for Astronomy, the University of Hawaii, the Pan-STARRS Project Office, the Max Planck Society and its participating institutes, the Max Planck Institute for Astronomy, Heidelberg, and the Max Planck Institute for Extraterrestrial Physics, Garching, The Johns Hopkins University, Durham University, the University of Edinburgh, the Queen’s University Belfast, the Harvard-Smithsonian Center for Astrophysics, the Las Cumbres Observatory Global Telescope Network Incorporated, the National Central University of Taiwan, the Space Telescope Science Institute, the National Aeronautics and Space Administration under grant No. NNX08AR22G issued through the Planetary Science Division of the NASA Science Mission Directorate, the National Science Foundation grant No. AST-1238877, the University of Maryland, Eotvos Lorand University (ELTE), the Los Alamos National Laboratory, and the Gordon and Betty Moore Foundation.
\end{acknowledgments}

\appendix

\section{Full corner plot}
\label{ap:full_corner_plot}

\begin{figure*}
\includegraphics[width=2.0\columnwidth]{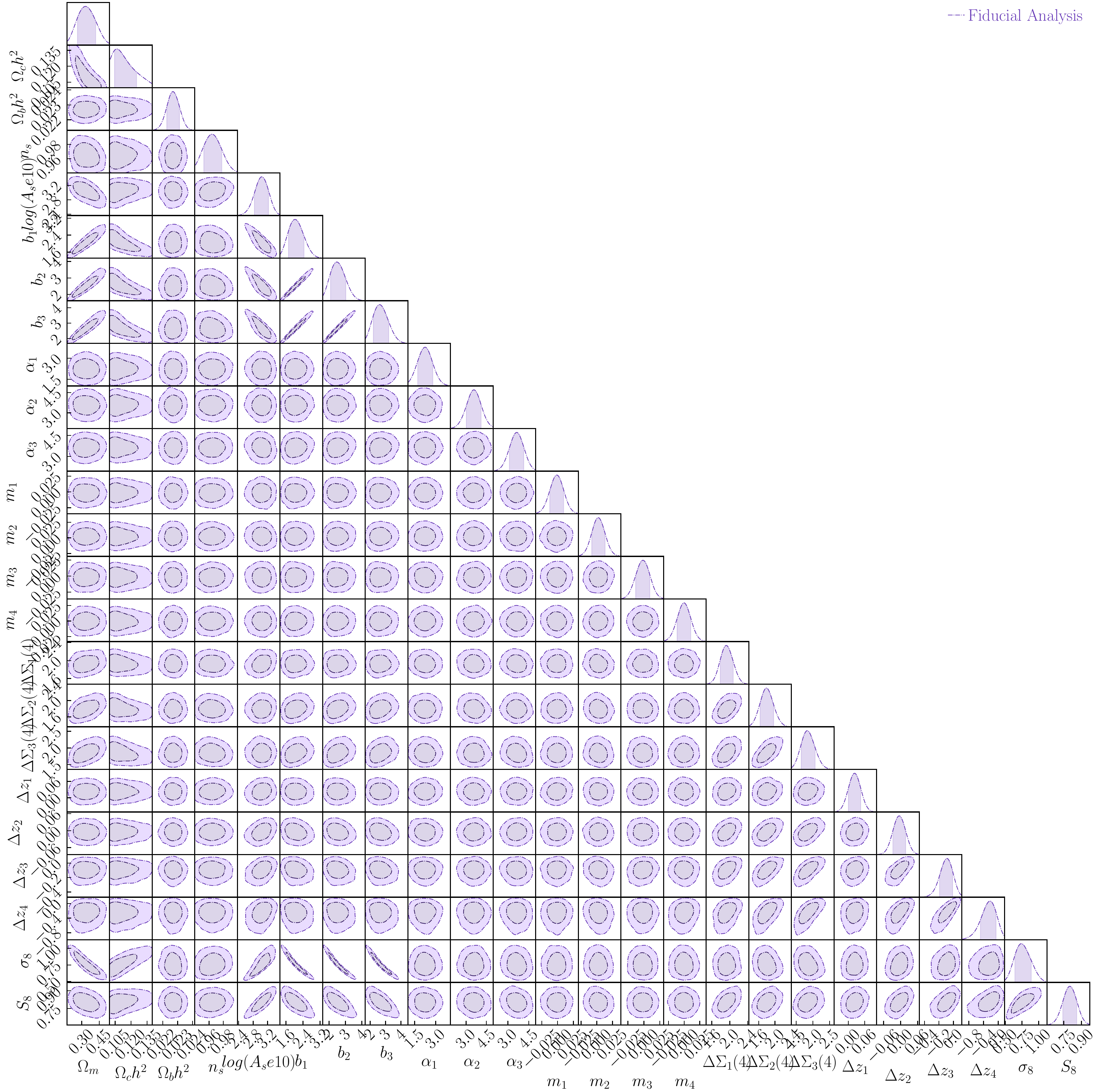}
\caption{\label{fig:full_corner_plot} The posterior distribution of the fiducial $2\times2$pt analysis for the 22 free parameters and two derived parameters ($\sigma_8$ and $S_8$).}
\end{figure*}

Fig~\ref{fig:full_corner_plot} shows the full corner plot of the 22 sampled parameters and two derived parameters ($\sigma_8$ and $S_8$). The parameter constraints are obtained through the fiducial $2\times2$pt analysis.

\bibliography{apssamp}

\end{document}